\documentclass[letterpaper,twocolumn,10pt]{article}
\usepackage{usenix-2020-09}
\usepackage[available,functional,reproduced]{usenixbadges}

\usepackage[utf8]{inputenc}
\usepackage{microtype}
\usepackage{graphicx,mathtools,kantlipsum}
\usepackage[normalem]{ulem}
\usepackage{url}
\usepackage{amsmath,amsfonts}
\usepackage{amsthm,enumitem}
\usepackage{listings}
\usepackage{courier}
\usepackage{float}
\usepackage{multirow}
\usepackage{makecell}
\usepackage{algorithm}
\usepackage[noend]{algpseudocode}
\usepackage{booktabs}
\usepackage[font=small]{caption}
\usepackage{subcaption}
\usepackage{hyperref}
\usepackage[capitalize]{cleveref}
\usepackage{balance}
\usepackage{pifont}
\usepackage{enumitem}
\usepackage{textcomp}
\usepackage{comment}
\usepackage{xcolor}

\usepackage[hang,flushmargin]{footmisc} 

\definecolor{codegreen}{rgb}{0,0.4,0}
\hypersetup{
    colorlinks=true,
    citecolor=blue,
    linkcolor=blue,
}

\usepackage{titlesec}
\titlespacing{\paragraph}{0pt}{5pt plus 3pt minus 1pt}{5pt plus 1pt minus 1pt}

\setlist[enumerate]{noitemsep,topsep=0pt,partopsep=3pt,leftmargin=*}
\setlist[itemize]{noitemsep,topsep=0pt,partopsep=3pt,leftmargin=*}

\usepackage{xspace}
\newcommand{\gpm}{{Graph Pattern Mining}\xspace}

\newcommand{\treemodel}{{search tree}\xspace}
\newcommand{\pv}{{pattern vertex}\xspace}
\newcommand{\dv}{{data vertex}\xspace}

\newcommand{\mo}{{matching order}\xspace}
\newcommand{\po}{{symmetry order}\xspace}

\newcommand{\dg}{{$\mathcal{G}$}\xspace}
\newcommand{\pg}{{$\mathcal{P}$}\xspace}
\newcommand{\vs}{{$\mathcal{V}$}\xspace}
\newcommand{\es}{{$\mathcal{E}$}\xspace}
\newcommand{\adj}{{$\mathcal{N}$}\xspace}
\newcommand{\sys}{{G\textsuperscript{2}Miner}\xspace}
\newcommand{\pbe}{{PBE}\xspace}

\crefformat{section}{\S#2#1#3}

\newcommand{\hl}[1]{\textcolor{black}{#1}}

\newcommand{\hlc}[1]{\textcolor{codegreen}{#1}}

\title{Efficient and Scalable \gpm on GPUs}
\author{{\rm Xuhao Chen} \\ MIT CSAIL \and {\rm Arvind} \\ MIT CSAIL}
\date{}
\begin{document}
\maketitle

\begin{abstract}
{\gpm (GPM) extracts higher-order information in a large graph by searching for small patterns of interest. GPM applications are computationally expensive, and thus attractive for GPU acceleration. Unfortunately, due to the complexity of GPM algorithms and parallel hardware, hand optimizing GPM applications suffers programming complexity, while existing GPM frameworks sacrifice efficiency for programmability. 
Moreover, little work has been done on GPU to scale GPM computation to large problem sizes.}

We describe \sys, the first GPM framework that runs efficiently on multiple GPUs.
\sys uses \textit{pattern-aware}, \textit{input-aware} and \textit{architecture-aware} search strategies to achieve high efficiency on GPUs. 
To simplify programming, it provides a code generator that automatically generates pattern-aware CUDA code.
\sys flexibly supports both breadth-first search (BFS) and depth-first search (DFS) to maximize memory utilization and generate sufficient parallelism for GPUs.
For the scalability of \sys, we propose a customized scheduling policy to balance workload among multiple GPUs.
Experiments on a V100 GPU show that \sys is 5.4$\times$ and 7.2$\times$ faster than the two state-of-the-art single-GPU systems, {Pangolin} and {PBE}, respectively. 
In the multi-GPU setting, \sys achieves linear speedups from 1 to 8 GPUs, for various patterns and data graphs.
We also show that \sys on a V100 GPU is 48.3$\times$ and 15.2$\times$ faster than the state-of-the-art CPU-based systems, Peregrine and GraphZero, on a 56-core CPU machine. 

\end{abstract}

\section{Introduction}

\begin{figure}[t]
\centering
\includegraphics[width=0.39\textwidth]{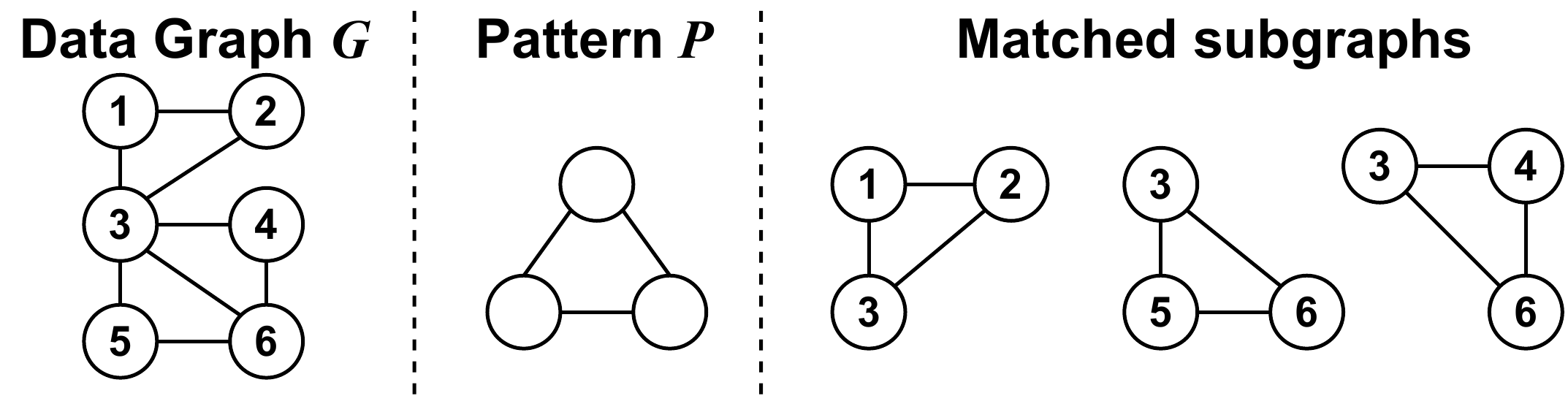}
\caption{\gpm example. The pattern \pg is 
		a triangle, and 3 triangles are found in the data graph \dg.}
\label{fig:gpm}
\end{figure}

\gpm (GPM) finds subgraphs in a given data graph which match the given pattern(s) (\cref{fig:gpm}).
GPM is a key building block in many domains,
e.g., protein function prediction~\cite{ppi,interactome1,Motif}, 
network alignment~\cite{milenkovic9,kuchaiev10},
spam detection~\cite{feldman11,becchetti13,domshlak}, 
chemoinformatics~\cite{ralaivola14,kashima15,chemical}, 
sociometric studies~\cite{holland4,Social},
image segmentation~\cite{zhang16}.
Graph machine learning tasks can also benefit from GPM, including anomaly detection~\cite{anomaly,anomaly-kdd}, 
entity resolution~\cite{entity}, 
community detection~\cite{graph-clustering}, role discovery~\cite{role} and relational classification~\cite{relational}.

GPM is extremely compute intensive, 
since it searches a space that is exponential in the pattern size.
For example, Peregrine~\cite{Peregrine}, a state-of-the-art GPM system on CPU, takes 9 hours to mine the {\tt 4-cycle} pattern (see \cref{fig:motifs}) in the {\tt Friendster} graph on a 56-core CPU machine.
GPUs provide much higher compute throughput and memory bandwidth than CPUs, and thus are attractive for GPM acceleration.

\begin{figure*}[thb]
\centering
\includegraphics[width=0.95\textwidth]{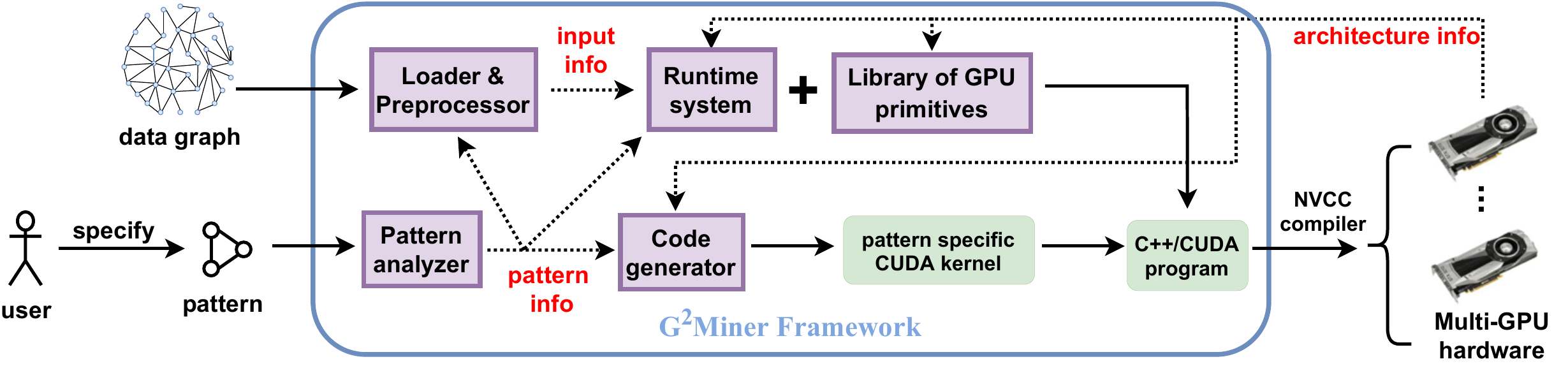}
\caption{\sys system overview. It contains a graph loader, a pattern analyzer, a runtime, a library of GPU primitives and a code generator.}
\label{fig:overview}
\end{figure*}


However, implementing GPM on GPU efficiently is challenging. This is because it requires sophisticated optimizations by leveraging information in the GPU hardware architecture, the pattern(s) of interest, and the input data graph.

\begin{itemize}

\item 
\emph{Architecture Awareness}: 
A GPU usually has smaller memory capacity than a CPU and requires more fine-grain data parallelism to be fully utilized. 
More threads, however, require more memory to accommodate intermediate data!
The search order, BFS or DFS, offers a similar tradeoff between memory and parallelism and therefore, GPM on GPU requires careful orchestration of parallelism and memory usage to maximize efficiency.
GPUs are also much more sensitive to thread divergence and workload imbalance~\cite{LonestarGPU} than CPUs.
This necessitates a  more sophisticated task-to-hardware mapping for GPU than that for CPU.

\item 
\emph{Pattern Awareness}: State-of-the-art GPM systems on CPU use \emph{pattern aware} search plans that prune the search space using pattern information.
This has been shown to be orders-of-magnitude faster than the pattern-oblivious search~\cite{Peregrine}.
{This pattern-aware approach has worked well for CPU, but it has not been well explored on GPU.
For example, many pattern-aware pruning schemes are only effective under DFS exploration, but existing GPU-based systems use BFS, and thus loss these opportunities for pruning}.

\item 
\emph{Input Awareness}:
Dynamic memory allocation is expensive in GPU~\cite{GPU-malloc} and we can avoid it if we can estimate the worst-case memory usage.
This can be done by using some meta information such as the maximum degree of the input graph.
For labelled graphs, we can use the vertex label distribution of the input graph to get the maximum number of possible patterns,
which helps save memory space.
In general, input information helps make better tradeoff among work efficiency, parallelism and memory consumption.
\end{itemize}

Given this complexity, the design goal of our GPM framework on GPUs involves the following considerations:

\begin{itemize}
\item 
\textbf{Efficiency:} 
To achieve high efficiency on GPU, a GPM system must be highly optimized with awareness of the pattern, the input and the hardware architecture.
{There is no prior system, neither on CPUs nor on GPUs, that considers all three aspects together.
This asks for a holistic solution that incorporates sophisticated optimizations systematically.} 

\item 
\textbf{Ease of programming:}
Writing efficient GPM code on GPUs is particularly difficult for domain users, who may not be parallel programming experts.
Thus, hiding GPU programming complexity is essential for system usability.

\item 
\textbf{Scalability:} 
The skewness in power-law graphs causes load imbalance.
This problem is exacerbated for DFS-based GPM algorithms,
because accesses to neighbors are multiple hops away. 
Hence, we need effective task scheduling and distribution policies to scale to multiple GPUs.
\end{itemize}

\begin{table}[b]
\centering
\resizebox{0.49\textwidth}{!}{
\begin{tabular}{c|c|c|c|c|c|c}
\Xhline{2\arrayrulewidth}
& \textbf{General} & \textbf{CPU} & \textbf{GPU} & \textbf{Multi-GPU} & \textbf{Order} & \textbf{Code Gen} \\ \hline
\textbf{EmptyHeaded~\cite{EmptyHeaded}} & & \checkmark  & & & DFS & \checkmark \\ \hline
\textbf{Graphflow~\cite{Graphflow,Delta-BigJoin,WCOJ}} & & \checkmark & & & DFS & \\ \hline
\textbf{GraphZero~\cite{AutoMine,GraphZero}} & & \checkmark & & & DFS & \checkmark \\ \hline
\textbf{GraphPi~\cite{GraphPi}} & & \checkmark & & & DFS & \checkmark\\ \hline
\textbf{Peregrine~\cite{Peregrine}} & \checkmark & \checkmark & & & DFS & \\ \hline
\textbf{Pangolin~\cite{Pangolin,Sandslash}} & \checkmark & \checkmark & \checkmark & & BFS &\\ \hline
\textbf{\pbe~\cite{GPU-Subgraph,reuse-subg}} & & & \checkmark & & BFS & \\ \hline
\textbf{\hl{\sys}} & \checkmark & & \checkmark & \checkmark & both & \checkmark \\
\Xhline{2\arrayrulewidth}
\end{tabular}
}
\caption{Comparison of state-of-the-art GPM systems, in terms of support for generality of the programming model, hardware platforms (CPU/GPU/multi-GPU), search orders, and code generation. }
\label{tab:sys-compare}
\end{table}

We propose \sys to overcome these challenges. 
\cref{tab:sys-compare} compares \sys to the state-of-the-art systems, including those that solve only the \textit{subgraph matching} problem, which is a subset of the GPM problem. 
In \cref{tab:sys-compare}, subgraph matching systems include EmptyHeaded, Graphflow, GraphZero, GraphPi and PBE, while Peregrine, Pangolin and \sys are general GPM systems. 
Much of the prior work focuses on CPU, and uses DFS to reduce the memory footprint.
GPU-based systems (Pangolin and PBE), on the other hand, use BFS because straightforward DFS implementations on GPU suffer from thread divergence and load imbalance.
This, however, limits their efficiency and/or the problem size they can solve.
Additionally, \sys simplifies GPU programming with automated CUDA code generation, 
while Pangolin requires users to write CUDA code manually, and PBE is not programmable at all.
Last but not least, \sys is the only system that scales to multiple GPUs. 

\cref{fig:overview} shows the overview of \sys.
It consists of a graph loader, a pattern analyzer, a runtime system, a library of CUDA primitives and a code generator.
The user is only responsible for specifying the pattern(s) of interest using our API (\cref{sect:overview}).
The pattern analyzer does analysis on the pattern and generates a \textit{pattern-specific} search plan,
based on which, the code generator (\cref{sect:codegen}) automatically generates pattern-specific CUDA kernels for GPUs. 
The kernels contain invocations to the device functions defined in the GPU primitive library (\cref{sect:primitives}) which includes efficiently implemented set operations.
The generated kernels, the GPU primitive library, and the runtime are compiled together by the NVCC compiler to generate the executable that runs on multi-GPU.

At runtime, the graph loader reads in the data graph, extracts input information (e,g., maximum degree and label distribution) and performs pattern-specific preprocessing on the data graph.
The pattern, input and architecture information is fed to the runtime (\cref{sect:runtime}) which heuristically handles GPU memory allocation, data transfer, and multi-GPU scheduling.


This paper makes the following contributions:

\begin{itemize}
\item \sys is the first pattern-aware, input data-graph-aware and architecture-aware framework for GPM, and it is the first GPM system that automates CUDA code generation for arbitrary patterns to simplify programming.

\item \sys is the first multi-GPU framework for GPM and the first GPU-based GPM framework that flexibly supports both BFS and DFS. 
It uses a novel task scheduling policy to balance workload among GPUs and we show \sys performance increases linearly from 1 to 8 V100 GPUs.

\item On a V100 GPU, \sys is 5.4$\times$ faster than \textit{Pangolin}, the only existing GPM system on GPU, and 7.2$\times$ faster than \textit{\pbe}, the state-of-the-art subgraph matching solver on GPU, thanks to the optimizations enabled in \sys (\cref{tab:optimizations}).

\item \sys on a V100 GPU is 48.3$\times$ and 15.2$\times$ faster than state-of-the-art CPU-based GPM system \textit{Peregrine} and subgraph matching system \textit{GraphZero} on a 56-core CPU.




\end{itemize}

\section{Background and Related Work}
\subsection{\gpm Problems}\label{subsect:define}


Let \dg(\vs, \es) be an undirected graph with \vs as the vertex and \es as the edge set. 
Given a vertex $v \in$ \vs , the neighbor set of $v$ is \adj($v$), 
the degree $d_v$ of $v$ is |\adj($v$)| and $\Delta$ is the maximum degree in \dg.
A graph $G'(W,F)$ is said to be a subgraph of \dg if $W \subseteq$ \vs and $F \subseteq$ \es.
$G'$ is a {\em vertex-induced subgraph} of \dg if $F$ contains all the edges in $E$ whose endpoints are in $W$.
$G'$ is an {\em edge-induced subgraph} of \dg if $W$ contains all the vertices in \vs which are the endpoints of edges in $F$.

\begin{figure}[t]
\centering
\includegraphics[width=0.43\textwidth]{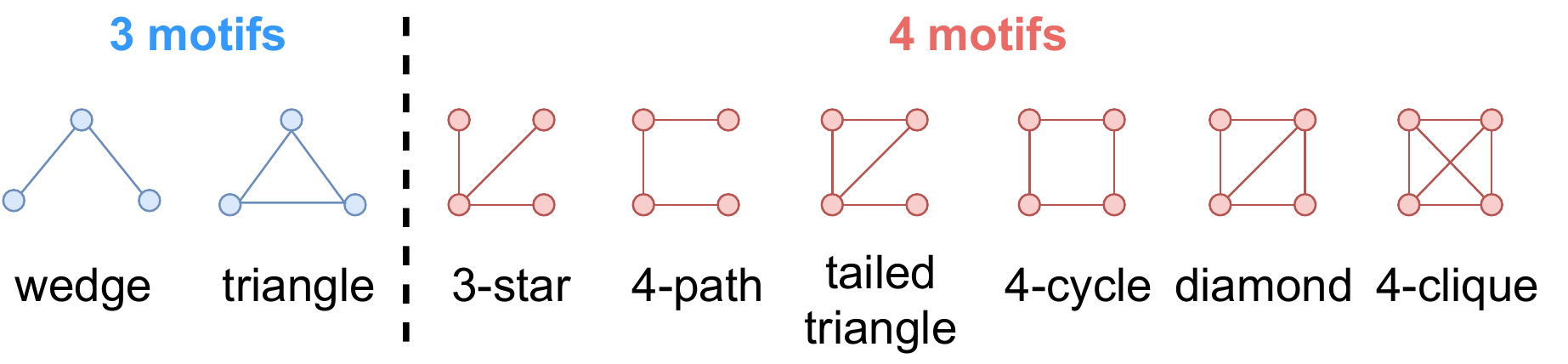}
\caption{3-vertex (left) and 4-vertex (right) motifs~\cite{Pangolin}.}
\label{fig:motifs}
\end{figure}

\textbf{Definition of GPM}. Given an undirected graph \dg and a set of patterns $S_p$=\{$P_1$, $P_2$, ...\} by the user, GPM finds vertex-induced or edge-induced subgraphs in \dg that are isomorphic to any \pg in $S_p$. 
If the cardinality of $S_p$ is 1, we call it a single-pattern problem. 
Otherwise, it is a multi-pattern problem.
The output of GPM varies in different problems, e.g., the pattern frequency (a.k.a, \textit{support}) or listing all matched subgraphs. 
The definition of support also varies, e.g., the count of matches or the \textit{domain support}~\cite{Pangolin} used in FSM.
Note that \textit{listing} requires enumerating every subgraph, but \textit{counting} does not. 
Thus, counting allows more aggressive search-space pruning. 

A pattern \pg is a small graph that can be defined explicitly or implicitly. An explicit definition specifies the vertices and edges of \pg, whereas an implicit definition specifies the desired properties of \pg.
For explicit-pattern problems, the solver finds matches of \pg in $S_p$. 
For implicit-pattern problems, $S_p$ is not known in advance. Therefore, the solver must find the patterns as well as their matches during the search.

GPM requires guarantee for {\em completeness}, i.e., every match of \pg in \dg should be found, and often {\em uniqueness}, i.e., every distinct match should be reported only once~\cite{Arabesque}.
To avoid confusion, we call a vertex in the pattern \pg as a {\em \pv} and denote it as $u_i$, and a vertex in the data graph \dg as a {\em \dv} and denote it as $v_i$.
Our work covers the following GPM problems from the literature~\cite{Pangolin,Fractal,Arabesque}:

\begin{itemize}
\item {\em Triangle counting} (TC): 
It counts the number of triangles (\cref{fig:gpm}), i.e., 3-cliques, in \dg.

\item {\em $k$-clique listing} ($k$-CL):
It lists all the $k$-cliques in \dg ($k \geq 3$). A $k$-clique is a $k$-vertex graph whose every pair of vertices are connected by an edge.

\item {\em Subgraph listing} (SL). It lists all edge-induced subgraphs of \dg that are isomorphic to a pattern \pg.

\item {\em $k$-motif counting} ($k$-MC):
It counts the number of occurrences of all possible $k$-vertex patterns.
Each pattern is called a {\em motif}~\cite{Motifs2,Motif3}. 
\cref{fig:motifs} shows all 3-motifs and 4-motifs.
This is also an example of a multi-pattern problem
because we have to find all the subgraphs that are isomorphic to \emph{any} pattern in a given set of patterns.

\item {\em $k$-frequent subgraph mining} ($k$-FSM): Given $k$ and a threshold $\sigma_{min}$, this problem considers all patterns with fewer than $k$ edges and lists a pattern \pg if the support $\sigma$ of \pg is greater than $\sigma_{min}$. This is called a {\em frequent} pattern. If $k$ is not specified, it is set to $\infty$, meaning that it is necessary to consider all possible values of $k$. 
In $k$-FSM, vertices in \dg have application-specific labels.
\end{itemize}

For TC and $k$-CL, vertex-induced and edge-induced subgraphs are the same. SL and FSM find edge-induced subgraphs, while $k$-MC looks for vertex-induced subgraphs. 
All problems seek to find explicit pattern(s) except FSM which finds implicit patterns.
$k$-MC and FSM are multi-pattern problems, while the others are single-pattern problems.

\begin{figure}[htb]
\centering
\includegraphics[width=0.49\textwidth]{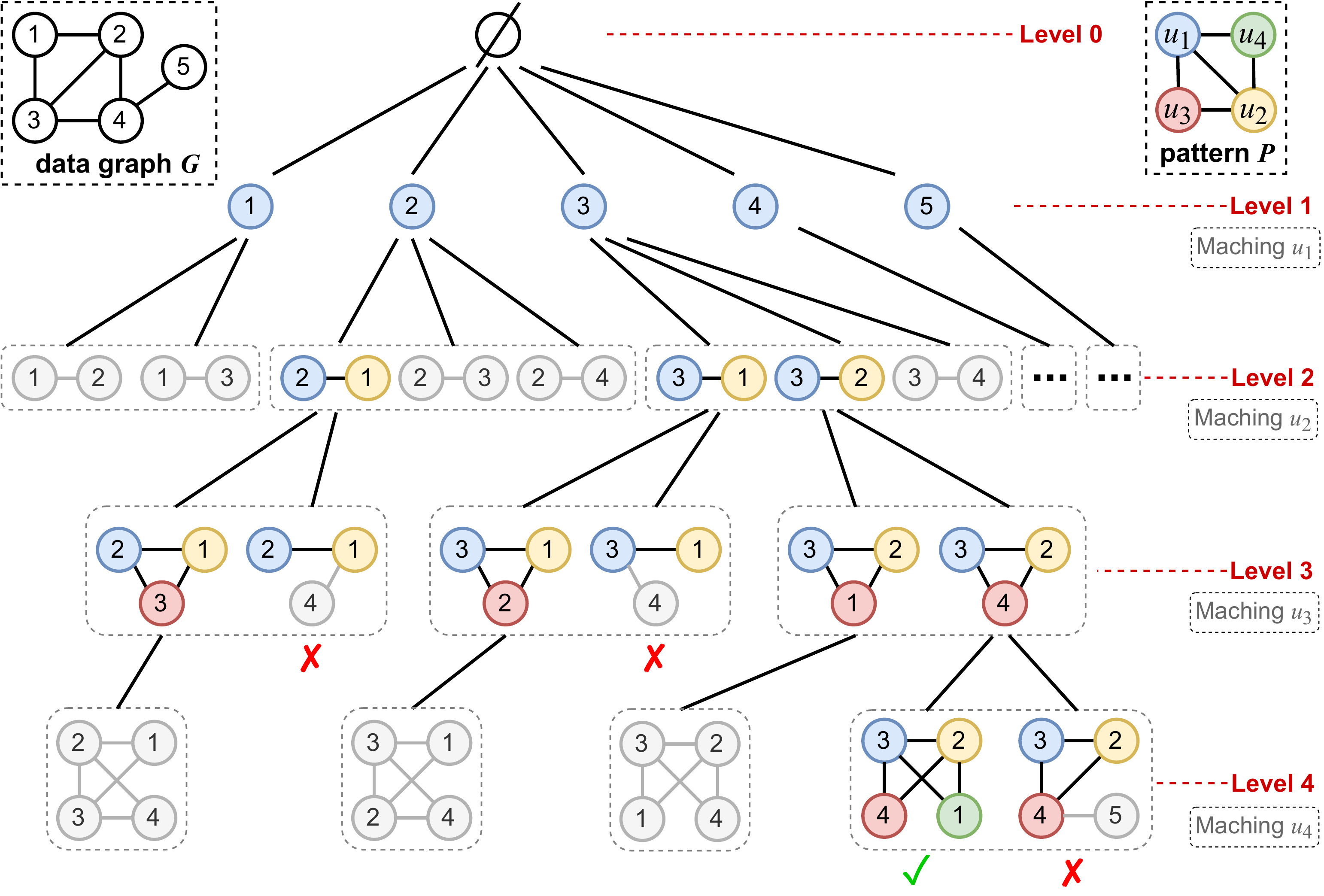}
\caption{A \treemodel using vertex extension. 
  Vertex colors (not vertex labels) show the matching between data vertices and pattern vertices.
  The matching order is \{$u_1 \rightarrow u_2 \rightarrow u_3 \rightarrow u_4$\}.
  The symmetry order is \{$v_a>v_b$, $v_c>v_d$\}. 
  Subgraphs in grey are ruled out by symmetry breaking.
  \textcolor{red}{$\times$} shows the unnecessary extensions that are pruned by the matching order. 
  \textcolor{green}{$\checkmark$} shows the matched subgraph. }
\label{fig:subgraph-tree}
\end{figure}

\begin{algorithm}[t]
\footnotesize
\caption{Pseudo code for finding {\tt diamond} in DFS order}
\label{alg:dfs-diamond}
\begin{algorithmic}[1]
  \For{{\bf each} vertex $v_1 \in$ \vs \hl{in parallel}}  \Comment{\hlc{match $v_1$ to $u_1$}} \label{algo:dfs:for1}
    \For{{\bf each} vertex $v_2 \in$ \adj($v_1$)} \Comment{\hlc{match $v_2$ to $u_2$}} \label{algo:dfs:for2}
	  \If{$v_2\ge v_1$} break; \Comment{\hlc{symmetry breaking}} \label{algo:dfs:break1}
	  \EndIf
	  \State $W \leftarrow$ \adj($v_1$) $\cap$ \adj($v_2$); \Comment{\hlc{set intersection: buffered in $W$}} \label{algo:dfs:buffer}
      \For{{\bf each} vertex $v_3 \in W$} \Comment{\hlc{match $v_3$ to $u_3$}} \label{algo:dfs:for3}
      \For{{\bf each} vertex $v_4 \in W$} \Comment{\hlc{match $v_4$ to $u_4$; $W$ is reused}} \label{algo:dfs:for4}
	      \If{$v_4\ge v_3$} break; \Comment{\hlc{symmetry breaking}} \label{algo:dfs:break2}
	      \Else $ $ count ++; \Comment{\hlc{do the counting}}
	      \EndIf
	    \EndFor
	  \EndFor
    \EndFor
  \EndFor
\end{algorithmic}
\end{algorithm}

\subsection{Pattern-Aware GPM Algorithms}
\label{subsect:solver}

A GPM problem is a search problem, whose search space is a {\it subgraph tree}~\cite{Sandslash,FlexMiner} (\cref{fig:subgraph-tree}). 
Each node in the tree is a subgraph of the data graph \dg. 
Subgraphs in level $l$ of the tree have $l$ vertices. 
The root of the tree (level 0) is an empty subgraph, while the leaves of the tree are potential candidates of matches.
A GPM problem can be solved by building this search tree,
and checking each leaf if it is isomorphic to the pattern \pg using the typical {\it graph isomorphism test}. 

The search tree is built by {\em vertex extension}:
subgraph $S_1{=}(W_1,E_1)$ can be extended by a single vertex $v \notin W_1$ to obtain subgraph $S_2{=}(W_2,E_2)$, 
if $v$ is connected to some vertex in $W_1$ (i.e., $v$ is in the {\em neighborhood} of subgraph $S_1$).  
When two subgraphs are related in this way, we say that $S_2$ is a child of $S_1$.
Formally, this can be expressed as $W_2 {=} W_1 \cup \{v\}$ where $v \notin W_1$ and there is an edge $(v,u) \in$ \es for some $u \in W_1$.
Similarly, \emph{edge extension} extends a subgraph $S_1$ with a single edge $(u,v)$, with at least one of the endpoints of the edge is in $S_1$.

The efficiency of a GPM algorithm depends heavily on how much we can prue the search tree. State-of-the-art GPM frameworks~\cite{AutoMine,Peregrine} use {\em pattern-aware} search plans 
that leverage the properties of the pattern to prune the tree. 
A pattern-aware search plan consists of a {\em \mo} and {\em \po}. 

\noindent
{\bf Matching order} is a total order that defines how the data vertices are matched to pattern vertices.
This order is used to eliminate irrelevant subgraphs on-the-fly.
As shown in \cref{fig:subgraph-tree}, to find the \texttt{diamond} pattern,
we use a matching order among pattern vertices: \{$u_1 \rightarrow u_2 \rightarrow u_3 \rightarrow u_4$\}, 
meaning that each vertex $v_1$ added at level 1 is matched to $u_1$;
each vertex $v_2$ added at level 2 are matched to $u_2$, and so on.
To search for matching candidates, there are connectivity constraints for the data vertices. 
For example, in {\tt diamond}, since $u_3$ is connected to both $u_1$ and $u_2$, candidate vertices of $v_3$ must be found in the intersection of $v_1$ and $v_2$'s neighborhoods, i.e., $v_3 \in$ \adj($v_1$) $\cap$ \adj($v_2$).
The same constraint should also be applied to $v_4$.
For a given pattern \pg, there exist multiple valid matching orders. 
To choose the best performing matching order,
prior works~\cite{AutoMine,Peregrine,GraphZero,GraphPi,Kudu,DwarvesGraph,DUALSIM,Delta-BigJoin} have proposed various cost models to predict the performance of matching orders,
and choose the one with the highest expected performance.

\noindent
{\bf Symmetry order}  is a partial order enforced among data vertices for \textit{symmetry breaking}, which removes redundant subgraph enumerations (a.k.a {\it automorphism}~\cite{Pangolin}), and thus guarantees that any match of \pg in \dg is found only once.
For example, for {\tt diamond}, we enforce that vertices added at level 1 must have larger ids than vertices added at level 2, i.e., $v_1>v_2$.
Thus, in level 2 of the tree in \cref{fig:subgraph-tree}, the subgraph \{2, 1\} is selected to be extended further, but subgraph \{1, 2\} is pruned. 
Similarly we add a constraint that $v_3>v_4$.
So the symmetry order for {\tt diamond} is \{$v_1>v_2$, $v_3>v_4$\}.

\subsection{DFS vs. BFS}

\begin{algorithm}[t]
\footnotesize
\caption{Pseudo code for finding Pattern \pg in BFS order}
\label{alg:bfs-diamond}
\begin{algorithmic}[1]
\For{{\bf each} level $i \in [1,$ \pg.size$]$}  \Comment{\hlc{level $i$ from 1 to the pattern size}} \label{algo:bfs:for1}
  \For{{\bf each} subgraph $sg \in SL_i$ \hl{in parallel}} \Comment{\hlc{$SL_i$: subgraph list}} \label{algo:bfs:for2}
    \For{{\bf each} vertex $u \in sg$} \label{algo:bfs:for3}
      \For{{\bf each} vertex $v \in$ \adj($u$)} \label{algo:bfs:for4}
        \State $sg'$ $\leftarrow sg \cup v$  \Comment{\hlc{vertex extension: add vertex $v$}} \label{algo:bfs:extend}
	    \If{$sg'$ satisfy \pg.constraints($i$)} 
	      \If{$i =$ \pg.size} count ++; \Comment{\hlc{leaf: a match found}}
	      \Else $ $  $SL_{i+1}$.insert($sg'$) \Comment{\hlc{go to the next level}} \label{algo:bfs:insert}
	      \EndIf
	    \EndIf
	  \EndFor
    \EndFor
  \EndFor
\EndFor
\end{algorithmic}
\end{algorithm}

Any search order (e.g., BFS, DFS) can be used to explore the search tree,
but different search orders come with different work efficiency, parallelism and memory consumption.

\begin{figure}[b]
\centering
\includegraphics[width=0.49\textwidth]{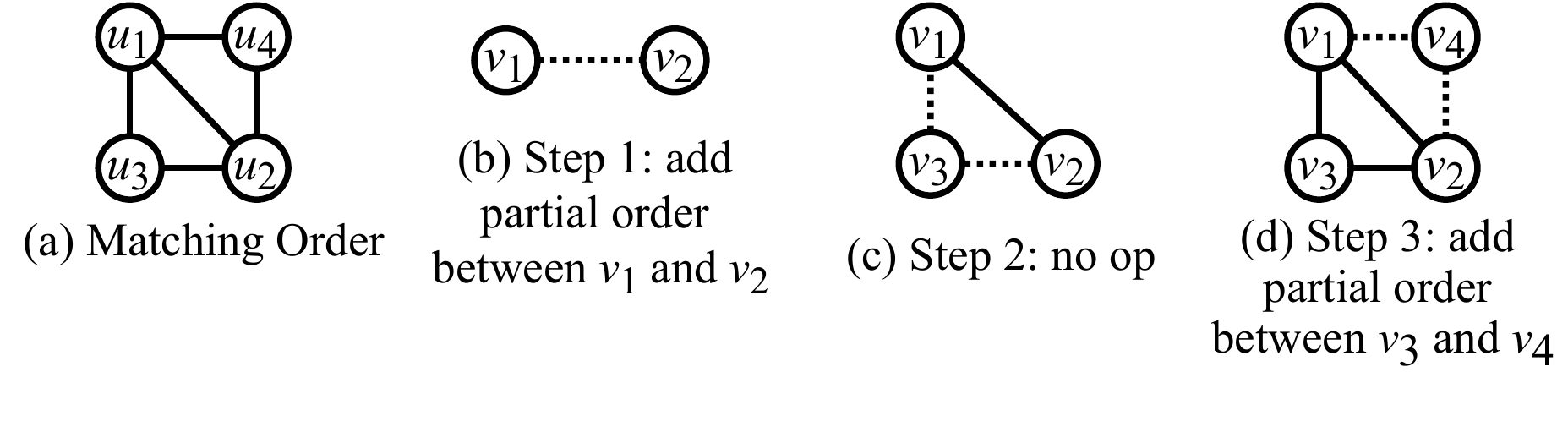}
\vspace{-20pt}
\caption{Generating \po for {\tt diamond} ~\cite{FlexMiner}.}
\label{fig:symmetry}
\end{figure}

\cref{alg:dfs-diamond} shows a DFS algorithm to mine the pattern {\tt diamond}.
It contains 4 nested {\tt for} loops (Line \ref{algo:dfs:for1}, \ref{algo:dfs:for2}, \ref{algo:dfs:for3}, \ref{algo:dfs:for4}).
Each loop corresponds to a data vertex ($v_1, v_2, v_3, v_4$) that is mapped to a pattern vertex ($u_1, u_2, u_3, u_4$) in \cref{fig:symmetry} (a).
A buffer $W$ in \cref{algo:dfs:buffer} holds intermediate data that is reused multiple times, which avoids redundant computation and thus improves \textit{work efficiency}.
The memory footprint contains only four vertices ($v_i$, $i=1,2,3,4$) and $W$ in \cref{algo:dfs:buffer} whose size is bounded by $\Delta$.
In DFS, every parallel \textit{task} does a DFS walk on the entire sub-tree rooted at $v_1$ (\cref{algo:dfs:for1}).
This is known as \textit{vertex parallelism}. The amount of parallelism is |\vs|.
Another way to parallelize it is \textit{edge parallelism},
in which every task contains the sub-tree rooted at each edge (say, if we make \cref{algo:dfs:for2} in parallel).
The amount of parallelism then is |\es|.

The BFS algorithm in \cref{alg:bfs-diamond} explores the tree level by level. 
In each level, it maintains a \textit{subgraph list} that is shared globally among all threads.
Each thread takes a subgraph from the subgraph list (\cref{algo:bfs:for2}),
and extends it to generate its child subgraphs (\cref{algo:bfs:extend}). 
The child subgraphs are inserted into the next-level subgraph list (\cref{algo:bfs:insert}).
In BFS, each parallel task is a subgraph in the subgraph list of the current level.
Since the size of the subgraph list increases exponentially level by level, 
the amount of parallelism increases rapidly. 
Although it provides more parallelism than DFS, BFS needs much more memory to accommodate the subgraph list.
For example, the BFS-based GPM system Pangolin~\cite{Pangolin} needs more than 40GB memory to mine the 5-{\tt clique} pattern in a moderate size graph {\tt livejournal},
making it impossible to run in most of off-the-shelf GPUs.

\subsection{GPM Systems and Applications}\label{sect:relate}

Many existing GPM systems \cite{Arabesque,RStream,Kaleido,G-Miner,Pangolin} use the BFS order.
{As they do level-by-level subgraph extension, they generate massive intermediate data and thus are limited to small graphs and patterns.}
Recently, a few DFS-based GPM systems \cite{Fractal,AutoMine,Sandslash,Peregrine} have been proposed to support larger datasets, {but they are all CPU-based}.
Among all, Pangolin is \textit{the only existing GPM system that supports GPU}. 
{However, limited by the BFS order, Pangolin can only handle small graphs, and it lacks pattern and input awareness.}

There also exist subgraph matching systems on CPU \cite{GraphZero,GraphPi,aDFS,Graphflow,EmptyHeaded} and GPU \cite{reuse-subg,GPU-Subgraph,Gsi}.
{But they only support a subset of GPM problems and are usually not programmable.}

Numerous hand-optimized GPM applications have been developed, including triangle counting \cite{Shun,TriCore,DistTC,GBBS,TC2017,TC2018,Suri,PDTL,TC2019,H-INDEX,comparative-tc}
, $k$-clique listing~\cite{KClique} and counting~\cite{Pivoter,ARB-COUNT,kclique-gpu,Counting-GPU}, 
motif counting \cite{PGD,ESCAPE,Rossi,Motif5-GPU,MotifGPU,CUDA-MEME},
subgraph listing/matching \cite{DUALSIM,Ullmann,PSgL,CECI,Cartesian,PathSampling,TurboFlux,Ma,Lai,Scaling,FRDSE,LIGHT,TurboIso,Matching-gunrock,Matching-GPU,Matching-GPU1,iso-gpu,Pgx-iso},
and FSM \cite{DistGraph,gSpan,GraMi,Scalemine,ParFSM,GpuFSM,FSM-GPU1,FSM-GPU2}.
{All of them are manually optimized with significant programming effort to achieve high efficiency,
which is quite a lot of burden for the domain programmers}.

\section{Challenges of Efficient GPM on GPU}
\label{sect:challenges}

\subsection{GPM vs. Graph Analytics}

Similar to graph analytics, GPM algorithms are \textit{irregular}~\cite{LonestarGPU} because the control flow and memory accesses are input-data dependent and thus, cannot be predicted statically. 
This irregularity causes random memory accesses and load imbalance, making it difficult to be efficiently parallelized.
Unlike graph analytics that only accesses 1-hop neighbors, 
GPM requires accessing multi-hop neighbors, 
which exacerbates the irregularity problem.
For example, load imbalance is much worse for DFS-based GPM than graph analytics because each parallel \textit{task} (a DFS walk on the entire sub-tree) is more coarse-grain in GPM.
In addition, GPM generates intermediate data during the search (buffer $W$ in DFS or subgraph list in BFS), which consumes extra memory than graph analytics.

\subsection{GPM on GPU vs. GPM on CPU} \label{sect:gpm-on-gpu}


Since the tasks are independent of each other, 
they are fairly easy to parallelize on CPU,
as shown in \cref{alg:dfs-diamond} \cref{algo:dfs:for1}.
But it is not as straightforward on GPU due to GPU's massively parallel model and limited memory capacity.

A GPU often consists of multiple \textit{streaming multiprocessors} (SM). 
Each SM accommodates multiple vector units.
This hardware organization results in a hierarchical parallel model:
each CUDA \textit{kernel} includes groups of threads called \textit{cooperative thread arrays} (CTAs) 
or \textit{thread blocks}.
Within each CTA, subgroups of threads called \textit{warps} are executed simultaneously.
Thus GPUs, to be fully utilized, require much more hierarchical parallelism than CPUs.

GPUs generally have less memory than CPUs,
while BFS-based GPM algorithms consume memory exponential in the pattern size.
Using DFS can reduce memory consumption, and also improve work efficiency.
Hence, state-of-the-art CPU-targeted GPM frameworks~\cite{AutoMine,GraphZero,Peregrine,Sandslash,GraphPi} all adopt DFS. 
However, naively porting the DFS-based CPU algorithms to GPU is not efficient because of the following reasons: 


{\bf (1) Branch Divergence.}
In \cref{alg:dfs-diamond}, each thread takes a vertex $v_1$ from \vs and starts DFS walk rooted by $v_1$.
Since different vertices have different neighborhoods, the threads in a warp may take different paths at the branches,
leading to inefficiency on GPU~\cite{DAWS}.
Branch divergence is much more severe for DFS than BFS due to the multiple nested loops for DFS backtracking that access multi-hop neighborhoods. 

{\bf (2) Memory Divergence.}
DFS walk also makes memory accesses more irregular.
This causes memory divergence in GPU, i.e., threads in a warp access non-consecutive memory locations.
In this case, each load instruction generates multiple
(up to the warp size, i.e., 32) memory requests to the memory subsystem,
which wastes memory bandwidth, congests on-chip data path~\cite{Chen-MICRO}, and thus results in poor GPU performance.

{\bf (3) Load Imbalance.}
Variance of neighborhood sizes in power-law graphs causes load imbalance.
In CPU it is less significant because there are limited number of cores/threads and each core is very powerful.
However, GPUs have thousands of lightweight cores and more than ten times the number of active threads.
If unbalanced, it would be much more costly since the slowest thread is running on a low-frequency core and thousands of cores are waiting.
Load imbalance is also less concerned for BFS, since it does level-by-level extension and at each level the tasks are lightweight, i.e., fine-grained. 

Therefore, existing GPU-based GPM systems~\cite{Pangolin} and subgraph matching systems~\cite{reuse-subg,GPU-Subgraph} all use BFS order.
This severely limits the graph sizes that they can handle.
PBE~\cite{GPU-Subgraph} partitions the data graph to support large graphs, but partitioning introduces cross-partition communication.
Note that using \textit{beam search}~\cite{beam-search} or bounded DFS does not fully resolve these issues, but loses the benefit of work efficiency of using DFS.

\section{\sys System Overview and Interface}\label{sect:overview}

We propose \sys (\cref{fig:overview}) to address the challenges in \cref{sect:challenges}. It hides away GPU programming complexity, and takes into account the properties of the pattern, input data graph and hardware architecture to achieve high efficiency on GPU. 
We first describe how to program in \sys in \cref{subsect:program}, 
and then introduce the system interface for extracting information out of the input, pattern and architecture (\cref{subsect:interface}).
Lastly we give an overview of the optimizations in \cref{subsect:optimizations}.

\begin{lstlisting}[
float=tp,floatplacement=t,
xleftmargin=0.35in,
linewidth=7.9cm,
label={lst:kcl},
basicstyle=\ttfamily\footnotesize,
language=C++, 
abovecaptionskip=0pt,
frame = single, 
firstnumber = last, 
escapeinside={(*@}{@*)},
escapeinside={(*}{*)},
escapechar=|,
morekeywords = {Vertex, Graph, Subgraph, Pattern, Set, Map, Match, void, int},
caption = {$k$-Clique Listing ($k$-CL) user code in \sys}]
Graph G = loadDataGraph("graph.csr");
Pattern p = generateClique(|$k$|); |\label{app:line:clique}|
list(G, p); // count(G, p) for counting
\end{lstlisting}

\begin{lstlisting}[
float=tp,floatplacement=t,
xleftmargin=0.35in,
linewidth=7.9cm,
label={lst:sl},
basicstyle=\ttfamily\footnotesize,
language=C++, 
abovecaptionskip=0pt,
frame = single, 
firstnumber = last, 
escapeinside={(*@}{@*)},
escapeinside={(*}{*)},
escapechar=|,
morekeywords = {Vertex, Graph, Subgraph, Pattern, Set, Map, Match, void, int},
caption = {Subgraph Listing (SL) user code in \sys}]
Pattern p("pattern.el", EdgeInduced);  |\label{app:line:sgl}|
list(G, p);
\end{lstlisting}

\subsection{Making Programming Easy}
\label{subsect:program}

\sys provides the same API as state-of-the-art CPU-based systems, 
e.g., Peregrine and Sandslash, making it friendly to users of CPU frameworks.
As shown in \cref{lst:kcl}, to program a $k$-CL solver in \sys,
the user specifies the pattern using an utility function {\tt generateClique()} (\cref{app:line:clique}),
and then call {\tt list()} to do listing or {\tt count()} to do counting.
If {\tt count()} is used, it allows the system enable counting-only optimizations (details in \cref{sect:codegen}).
To list an arbitrary pattern \pg (\cref{lst:sl}), the user can specify \pg using its edgelist ({\tt pattern.el} at \cref{app:line:sgl}).
By default \sys finds vertex-induced subgraphs. 
Since SL requires listing edge-induced subgraphs by definition,
the user needs to specify it (\texttt{EdgeInduced} at \cref{app:line:sgl}).

\begin{lstlisting}[
float=tp,floatplacement=t,
xleftmargin=0.2in,
linewidth=8.3cm,
label={lst:kmc},
basicstyle=\ttfamily\footnotesize,
language=C++, 
abovecaptionskip=0pt,
frame = single, 
firstnumber = last, 
escapeinside={(*@}{@*)},
escapeinside={(*}{*)},
escapechar=|,
morekeywords = {Vertex, Graph, Subgraph, Pattern, Set, Map, Match, void, int},
caption = {$k$-Motif Counting ($k$-MC) user code in \sys}]
Set<Pattern> patterns = generateAll(|$k$|); |\label{app:line:motif}|
Map<Pattern,int> result = count(G, patterns);
\end{lstlisting}

\begin{lstlisting}[
float=tp,floatplacement=t,
xleftmargin=0.17in,
linewidth=8.3cm,
label={lst:fsm},
basicstyle=\ttfamily\footnotesize,
language=C++, 
abovecaptionskip=0pt,
frame = single, 
firstnumber = last, 
escapeinside={(*@}{@*)},
escapeinside={(*}{*)},
escapechar=|,
morekeywords = {Vertex, Graph, Subgraph, Pattern, Set, Map, Match, Domain, void, int},
caption = {Frequent Subgraph Mining ($k$-FSM) user code in \sys}]
Void updateSupport(Subgraph s) { |\label{app:line:support}|
  map(s.getPattern(), s.getDomain());
}
bool patternFilter(Pattern p) { |\label{app:line:filter}|
  return p.getDomainSupport() >= threshold;
}
Set<Pattern> patterns = generateAll(|$k$|,   
                 EdgeInduced, patternFilter);
list(G, patterns, PATTERN_ONLY); |\label{app:line:fsm}|
\end{lstlisting}

For multi-pattern problems, the user is interested in a set of patterns instead of just one.
For $k$-MC in \cref{lst:kmc}, the patterns can be generated by calling an utility function {\tt generateAll()} (\cref{app:line:motif}) or parsing the patterns' edgelists.

Programmability is particularly important for implicit-pattern problems. 
The user must implement API functions to specify the patterns.
For example, for $k$-FSM in \cref{lst:fsm}, the user chooses to use \texttt{domain support} by implementing \texttt{updateSupport} (\cref{app:line:support}).
To specify the properties that differentiate the interesting patterns with irrelevant patterns, the user must define \texttt{patternFilter} (\cref{app:line:filter}).
As FSM asks for only listing the patterns, we can specify a \texttt{PATTERN\_ONLY} keyword in \texttt{list} to avoid listing the subgraphs (\cref{app:line:fsm}). 
If the user wants to customize the output, one can define a {\tt output()} function and pass it to \texttt{list}, instead of using \texttt{PATTERN\_ONLY}. 
This function defines custom operations on each subgraph of interest, 
which can also be used to do \textit{early termination}~\cite{Peregrine} by checking a user-defined condition.

\subsection{System Interface} \label{subsect:interface}

The pattern specified by user API is fed to a \textit{pattern analyzer} to extract useful pattern information.
Meanwhile, the GPU hardware information is taken by \sys to enable optimizations in the runtime, code generator and GPU primitives.
At runtime, the data graph is loaded by a \textit{graph loader} which collects input information and also performs preprocessing.

\noindent
\textbf{Pattern Analyzer}.
The \textit{pattern analyzer} generates: (1) a search plan with a \mo and a \po, which is used by the code generator; (2) reuse opportunities using buffers (e.g., $W$ in \cref{alg:dfs-diamond}), used by the code generator and the runtime; (3) other important properties of the pattern, e.g., whether the pattern is a clique or \texttt{hub-pattern} (\cref{subsect:pruning} (2)), used by the runtime and code generator.

The pattern analyzer enumerates all the possible \mo{s} of \pg, and uses a cost model to pick the best one. 
We use the same cost model as GraphZero~\cite{GraphZero} for fair comparison, but any cost model can be employed by \sys.
We also use the algorithm in GraphZero to generate a \po: it takes the generated \mo $\mathcal{MO}$
and builds a subgraph incrementally in the order specified by $\mathcal{MO}$.
At each step it detects symmetric vertex pairs and adds orders accordingly. 
For example, for \texttt{diamond},
the matching order in \cref{fig:symmetry} (a) results in the three steps shown in (b), (c) and (d),
during which we add partial order $v_2<v_1$ and $v_4<v_3$.


\begin{table*}[t]
\resizebox{\textwidth}{!}{%
\begin{tabular}{cl|ccccc|c|c|c}
\Xhline{2\arrayrulewidth}
\multicolumn{2}{c|}{} & \multicolumn{5}{c|}{\textbf{Effect}} & & & \\ \cline{3-7}
\multicolumn{2}{c|}{\multirow{-2}{*}{\textbf{Optimizations}}} & \multicolumn{1}{c|}{\begin{tabular}[c]{@{}c@{}}mitigate\\ divergence\end{tabular}} & \multicolumn{1}{c|}{\begin{tabular}[c]{@{}c@{}}load\\ balance\end{tabular}} & \multicolumn{1}{c|}{\begin{tabular}[c]{@{}c@{}}mem.\\ saving\end{tabular}} & \multicolumn{1}{c|}{\begin{tabular}[c]{@{}c@{}}algorithm\\ pruning\end{tabular}} & \begin{tabular}[c]{@{}c@{}}extra GPU\\ efficiency\end{tabular} & \multirow{-2}{*}{\textbf{\begin{tabular}[c]{@{}c@{}}Used in\\ Pangolin?\end{tabular}}} & \multirow{-2}{*}{\textbf{\begin{tabular}[c]{@{}c@{}}Used in hand \\ written apps?\end{tabular}}} & \multirow{-2}{*}{\textbf{Conditions to apply}} \\ \hline
\multicolumn{1}{c|}{\begin{tabular}[c]{@{}c@{}}\textbf{Category-(1)}:\\ Known\end{tabular}} & \begin{tabular}[c]{@{}c@{}}A: Data graph preprocessing\\ (edge orientation)~\cref{subsect:interface}\end{tabular} & \multicolumn{1}{c|}{} & \multicolumn{1}{c|}{} & \multicolumn{1}{c|}{\checkmark} & \multicolumn{1}{c|}{\checkmark} & & \checkmark & \checkmark & cliques \\ \hline
\multicolumn{1}{c|}{} & \begin{tabular}[c]{@{}c@{}}B: Data graph partitioning \\ \cref{sect:mem-mgmt} (1)\end{tabular} & \multicolumn{1}{c|}{} & \multicolumn{1}{c|}{} & \multicolumn{1}{c|}{\checkmark} & \multicolumn{1}{c|}{} & & $\times$ & TC only & \begin{tabular}[c]{@{}c@{}}hub patterns, graph size, \\ GPU memory size\end{tabular} \\ \cline{2-10} 
\multicolumn{1}{c|}{} & \begin{tabular}[c]{@{}c@{}}C: Two-level parallelism \\ ~\cref{sect:parallelism}\end{tabular} & \multicolumn{1}{c|}{\checkmark} & \multicolumn{1}{c|}{\checkmark} & \multicolumn{1}{c|}{} & \multicolumn{1}{c|}{} & & $\times$ & TC only & always enabled on GPU \\ \cline{2-10} 
\multicolumn{1}{c|}{} & \begin{tabular}[c]{@{}c@{}}D: Counting-only pruning \\ \cref{subsect:pruning} (1)\end{tabular} & \multicolumn{1}{c|}{} & \multicolumn{1}{c|}{} & \multicolumn{1}{c|}{} & \multicolumn{1}{c|}{\checkmark} & & $\times$ & CPU only & \begin{tabular}[c]{@{}c@{}}automatic pattern \\ decomposition~\cite{ESCAPE}\end{tabular} \\ \cline{2-10} 
\multicolumn{1}{c|}{} & \begin{tabular}[c]{@{}c@{}}E: Local graph search \\ \cref{subsect:pruning} (2)\end{tabular} & \multicolumn{1}{c|}{} & \multicolumn{1}{c|}{} & \multicolumn{1}{c|}{} & \multicolumn{1}{c|}{\checkmark} & & $\times$ & CL only &  \\ \cline{2-9}
\multicolumn{1}{c|}{} & \begin{tabular}[c]{@{}c@{}}F: Flexible data format \\ \cref{subsect:flexible-data}\end{tabular}  & \multicolumn{1}{c|}{\checkmark} & \multicolumn{1}{c|}{} & \multicolumn{1}{c|}{} & \multicolumn{1}{c|}{} & \multicolumn{1}{c|}{} & $\times$ & CC only & \multirow{-2}{*}{\begin{tabular}[c]{@{}c@{}}hub patterns \&\\ $\Delta$ \textless 1024\end{tabular}} \\ \cline{2-10} 
\multicolumn{1}{c|}{\multirow{-10}{*}{\begin{tabular}[c]{@{}c@{}}\textbf{Category-(2)}:\\ \\ Known, but not \\ enabled in prior \\ GPM systems\end{tabular}}} & \begin{tabular}[c]{@{}c@{}}G: Multi-gpu scheduling \\ ~\cref{sect:scheduler}\end{tabular} & \multicolumn{1}{c|}{} & \multicolumn{1}{c|}{\checkmark} & \multicolumn{1}{c|}{\textbf{}} & \multicolumn{1}{c|}{} & & $\times$ & MC only  & always used on multi-GPU \\ \hline
\multicolumn{1}{c|}{} & \begin{tabular}[c]{@{}c@{}}H: SIMD-aware primitives \\ \cref{subsect:simd}\end{tabular} & \multicolumn{1}{c|}{\textbf{}}  & \multicolumn{1}{c|}{\textbf{}} & \multicolumn{1}{c|}{\textbf{}} & \multicolumn{1}{c|}{} & \checkmark & $\times$ & $\times$ & \begin{tabular}[c]{@{}c@{}}hardware support for \\ warp level primitives\end{tabular}   \\ \cline{2-10} 
\multicolumn{1}{c|}{} & \begin{tabular}[c]{@{}c@{}}I: Multi-pattern fission \\ ~\cref{subsect:multi-pattern}\end{tabular} & \multicolumn{1}{c|}{} & \multicolumn{1}{c|}{} & \multicolumn{1}{c|}{} & \multicolumn{1}{c|}{} & \checkmark & $\times$ & $\times$ & \begin{tabular}[c]{@{}c@{}}explicit multi-pattern \& \\ kernel occupancy by NVCC\end{tabular}    \\ \cline{2-10}
\multicolumn{1}{c|}{} & \begin{tabular}[c]{@{}c@{}}J: Edgelist reduction \\ ~\cref{sect:mem-mgmt} (2)\end{tabular} & \multicolumn{1}{c|}{} & \multicolumn{1}{c|}{} & \multicolumn{1}{c|}{\checkmark} & \multicolumn{1}{c|}{} &  & $\times$ & $\times$ & if $v_0$ \textgreater $v_1$ in symmetry order \\ \cline{2-10} 
\multicolumn{1}{c|}{} & \begin{tabular}[c]{@{}c@{}}K: Adaptive buffering \\ ~\cref{sect:mem-mgmt} (3)\end{tabular} & \multicolumn{1}{c|}{} & \multicolumn{1}{c|}{} & \multicolumn{1}{c|}{\checkmark} & \multicolumn{1}{c|}{} & & $\times$ & $\times$ & \begin{tabular}[c]{@{}c@{}}buffer $W$ usage in matching \\ order \& GPU memory size\end{tabular}   \\ \cline{2-10} 
\multicolumn{1}{c|}{} & \begin{tabular}[c]{@{}c@{}}M: Hybrid order on GPU \\ \cref{sect:orders}\end{tabular} & \multicolumn{1}{c|}{} & \multicolumn{1}{c|}{} & \multicolumn{1}{c|}{\checkmark} & \multicolumn{1}{c|}{} & & $\times$ & $\times$ & \begin{tabular}[c]{@{}c@{}}implicit, intermediate data \\ unbounded, user-specified\end{tabular} \\ \cline{2-10} 
\multicolumn{1}{c|}{\multirow{-11}{*}{\begin{tabular}[c]{@{}c@{}}\textbf{Category-(3)}:\\ \\ Novel \\ for GPM\end{tabular}}} & \begin{tabular}[c]{@{}c@{}}N: memory reduction using \\ \ \ \ \ label frequency ~\cref{sect:mem-mgmt} (4)\end{tabular} & \multicolumn{1}{c|}{} & \multicolumn{1}{c|}{} & \multicolumn{1}{c|}{\checkmark} & \multicolumn{1}{c|}{} & & $\times$ & $\times$ & \begin{tabular}[c]{@{}c@{}}implicit, vertex label \\ frequency, user-specified\end{tabular} \\ \Xhline{2\arrayrulewidth}
\end{tabular}%
}
\caption{{Optimizations in G2Miner. Among them, optimizations A, B, D, E, F, I, J, K, M, N are pattern-aware; 
optimizations B, C, G, H, I, K, M are architecture-aware; 
and optimizations B, E, F, K, N are input-aware.
Pattern-aware optimizations are applied based on the pattern analysis, while input-aware and architecture-aware optimizations are enabled according to the input and architecture information, respectively. 
TC: triangle counting. CL/CC: clique listing/counting. MC: motif counting.}}
\label{tab:optimizations}
\end{table*}

\noindent
\textbf{Graph Loader and Preprocessor}.
The data graph \dg is loaded by the \textit{graph loader} into the memory in the compressed sparse row (CSR) format.
As \dg is being loaded, useful input information of the data graph is extracted,
e.g., |\vs|, |\es| and $\Delta$ of \dg.
In addition, if the graph is labelled, the vertex frequency of each label is computed (see usage for FSM in~\cref{sect:mem-mgmt}).
After \dg is loaded into memory, some preprocessing is performed on \dg.
First, the neighbor list of each vertex is sorted by ascending order of vertex IDs, so that we can apply early exit when we search the list with an upper bound (i.e., symmetry breaking).
Second, if a pattern of \texttt{clique} is detected, \sys enables a typical optimization called \textit{orientation}~\cite{Pangolin} .
{It gives every edge a direction in the undirected data graph \dg, 
which in turn converts \dg into a directed graph.
This halves the edge count in \dg, 
significantly reduces $\Delta$, and completely eliminates on-the-fly checking.}
Third, our preprocessor also supports sorting (e.g., by degree) and renaming the vertices in \dg to improve load balance \cite{Peregrine,GraphZero}.
Note that all these preprocessing operations need to be done only once.

\subsection{Overview of Optimizations}
\label{subsect:optimizations}

\cref{tab:optimizations} lists all the optimizations enabled in \sys.
\hl{We classify them into three categories.
Optimizations in Category-(1) are those exist in prior GPM systems.
Optimizations in Category-(2) do not exist in prior GPM systems (e.g., Pangolin) but have been used in some hand-written GPM applications. 
For example, optimization \texttt{D: data graph partitioning} has only been used for triangle counting, 
while in \sys we generalize it for all the clique patterns.
These optimizations are missing in prior GPM systems because prior systems are oblivious to the required pattern, input or architecture information. 
Optimizations in Category-(3) are novel as they have never been used for GPM, 
though some of them are known for GPU computing in general.}

\hl{As shown in column 3 to 7 of \cref{tab:optimizations}, these optimizations have different kinds of effect on GPM applications: 
(1) mitigating thread divergence; 
(2) improving load balancing; 
(3) reducing memory consumption; 
(4) pruning search space; 
and (5) improving efficiency based on GPU hardware features.}

\hl{The last column of \cref{tab:optimizations} shows the conditions for each optimization to be applied.
All the optimizations in \cref{tab:optimizations} are automated in \sys based on detecting the conditions, except for M and N (the last two rows). 
M and N are particularly used for implicit-pattern problems like FSM, for which the system cannot infer the conditions automatically. Thus, M and N are user-activated by specifying a flag.
}

Next, we describe these optimizations in detail, in the three major components of \sys: 
the code generator (\cref{sect:codegen}), the device function library (\cref{sect:primitives}) and the runtime scheduler (\cref{sect:runtime}).

\section{Pattern-specific GPU Code Generation} \label{sect:codegen}

\sys includes a \textit{pattern-aware} code generator that automatically generate CUDA code specific to the pattern. 
Prior work~\cite{AutoMine,GraphZero} has explored how to generate pattern-specific CPU code based on the \mo and \po, but code generation is more challenging for GPU.

Generating pattern-specific CPU code is relatively straightforward.
For example, to generate \cref{alg:dfs-diamond} for {\tt diamond}, 
the matching order in \cref{fig:symmetry} (a) is used to generate the 4 nested {\tt for} loops,
and the symmetry order is then used to insert breaks at Line \ref{algo:dfs:break1} and \ref{algo:dfs:break2}.
Whenever a set operation is needed,
a function call to the set operation primitive (implemented in a library) is inserted (\cref{algo:dfs:buffer}).
Since $v_3$ and $v_4$ are both from \adj($v_1$) $\cap$ \adj($v_2$),
a buffer $W$ is created for data reuse.
Finally, {\it task parallelism} is used to parallelize the program,
i.e., each thread processes one task at a time (\cref{algo:dfs:for1}).

However, generating efficient GPU code is more challenging, because 
(1) DFS-based GPM suffers from the thread divergence and load imbalance issues (\cref{sect:parallelism}); 
(2) hybrid search orders are needed in some cases (\cref{sect:orders}); and
(3) extra support is needed for 
 multi-pattern problems (\cref{subsect:multi-pattern}) and advanced pruning schemes (\cref{subsect:pruning}).

\subsection{Parallel Strategies for DFS on GPU} \label{sect:parallelism}
To maximize GPU efficiency for the DFS algorithm, we employ a \textit{two-level parallelism} strategy in \sys to exploit both inter-warp task parallelism and intra-warp \textit{data parallelism}.
This is motivated by our key observation that in GPM algorithms \textit{most of execution time is spent on set operations}. 
For example, when we executed Peregrine on a multicore CPU, set operations for each benchmark took 75\% to 92\% of the total execution time.
This motivated us to parallelize set operations by exploiting the data parallelism within each warp.
It alleviates divergence and also provides more parallelism to fully utilize GPUs.
To reduce load imbalance and further increase parallelism, we use {edge parallelism} for GPU instead of the {vertex parallelism} used for CPU.

\noindent
\textbf{(1) Reduce divergence with warp-centric parallelism.}
We could map each task to a thread, a warp or a CTA in a GPU.
As DFS has much more coarse-grained tasks than BFS,
mapping a task to a thread would be highly divergent and unbalanced for GPUs.
However, if we map a task to a CTA, 
all (e.g., 256) threads in the CTA will be used to process the same set operations.
If the two input neighbor-lists of a set operation are small, many threads in the CTA will be idle, leading to low utilization.
Moreover, all threads in the CTA will do the same DFS walk, which is a lot of redundant computation.

In \sys we use \textit{warp-centric data parallelism}.
Each task is assigned to a warp.
All threads in a warp synchronously perform the same DFS walk of the task.
During the DFS walk, whenever a set operation is encountered,
all threads in the warp work cooperatively to compute the set operation in parallel.
It has several benefits. 
First it achieves higher throughput than CPU since set operations are parallelized.
Second, it alleviates thread divergence within each warp as all threads in a warp are progressing synchronously. 
Third, it causes less redundancy than using CTA.
Our evaluation shows it is on average 2$\times$ faster than CTA-centric parallelism.

\begin{figure}[t]
\centering
\includegraphics[width=0.49\textwidth]{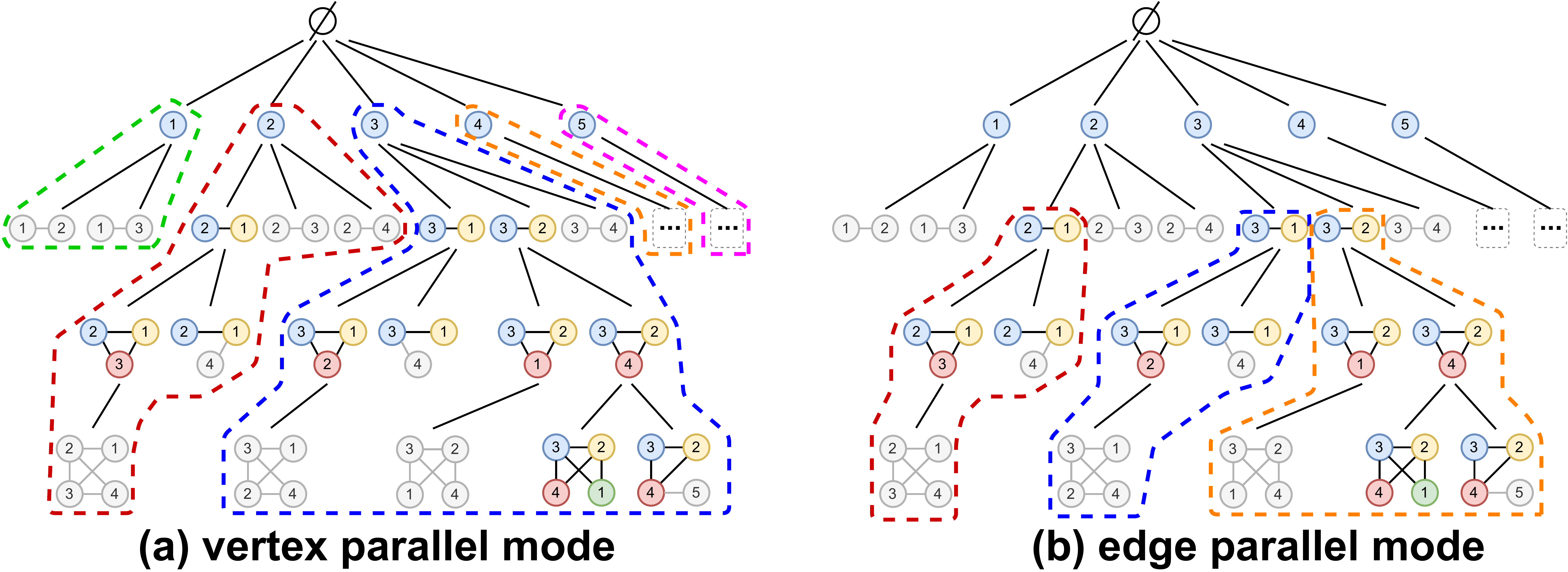}
\caption{(a) vertex-parallel vs. (b) edge-parallel execution. Each dashed circle is a parallel task. 
A task is mapped to one thread on CPU, but in \sys it is mapped to one warp on GPU.}
\label{fig:vertex-edge}
\end{figure}

\noindent
{\bf (2) Reduce task granularity for load balance.}
GPM systems on CPU use vertex parallelism~\cite{Peregrine,Sandslash,AutoMine,GraphZero}, i.e., each task is a DFS walk rootedin a vertex, as shown in \cref{fig:vertex-edge} (a).
This can already provide enough parallelism for CPU, needs no auxiliary data,
and potentially enjoys data reuse within the sub-tree.
But the coarse-grain tasks lead to load imbalance which can not be well tolerated by GPUs.
To reduce task granularity, we use {edge-parallelism}, i.e.,
each task explores the subtree rooted by an edge. 
As shown in \cref{fig:vertex-edge} (b),
apparently more work is required to search the subtree below a vertex on average compared to searching the subtree below an edge.
In addition to better load balance, edge parallelism can provide more parallelism (|\es|>|\vs|) for GPU than vertex parallelism.

\begin{itemize}
\item
By default, our code generator generates edge-parallel kernels.
Our evaluation shows they are mostly (1.5$\times$ on average) faster than vertex parallel ones.
But some GPM algorithms must use vertex parallelism.
For example, the 3-MC algorithm in \cite{Sandslash} can only be done in vertex parallelism.
\sys supports both vertex and edge parallelism. 
The user can set a compiler flag to use vertex parallelism, in which case $\Omega$ is not generated to save memory.



\end{itemize}

\noindent
\hl{\textbf{Discussion}. \textit{Two-level parallelism} has been only used for triangle counting~\cite{TriCore}, and it is challenging to extend it for all GPM problems.
First, triangle counting does subgraph extension only once, which needs no DFS traversal. 
Thus, \sys is the first to support DFS for GPM on GPU. 
Second, naive GPU implementations for complex patterns can easily run out of memory for intermediate data. 
This is not a concern for triangle counting. 
Third, during the DFS traversal, it requires extension to support high-performance generic set operations and multi-pattern, 
which triangle counting does not require. 
In the following, we show that these challenges can be resolved by applying optimizations H, I, J, K, M, N in \cref{tab:optimizations}.}

\subsection{Support for Hybrid Search Orders} \label{sect:orders}

With the two-level parallelism in \cref{sect:parallelism}, for many GPM problems, DFS is faster than BFS in \sys. 
However, this is not the case for problems like FSM. 
FSM computes the \texttt{domain support} and thus requires aggregating all the subgraphs for each pattern to compute its support.
In the DFS-based FSM algorithm~\cite{DistGraph,gSpan}, each task is a single-edge pattern (instead of subgraph) and the entire subtree of that pattern. 
This is \textit{pattern-parallel}, instead of vertex-parallel or edge-parallel.
Since the number of patterns is much smaller than the number of vertices or edges, 
the parallelism in FSM is not sufficient for GPU.
Moreover, the task granularity in pattern-parallelism is much larger than that in vertex- or edge-parallelism,
making the problem even more unbalanced.


In \sys we use a hybrid of BFS and DFS, or \textit{bounded BFS} search for problems that use \texttt{domain support} (e.g., FSM). 
At the single-edge level (i.e., level-2), we start with BFS search to aggregate edges by their patterns in parallel, which provides abundant parallelism.
As the search goes deeper, the number of subgraphs increases exponentially. 
To fit the intermediate data in memory, we divide the subgraphs into blocks.
Each block has a size that can resides in GPU memory, but also contains enough amount of subgraphs that can fully utilize the GPU.
Once the current block is processed, it moves to the next block.
Using this bounded BFS search, \sys can support larger graphs than Pangolin.

\subsection{Support for Multi-pattern Problems}  \label{subsect:multi-pattern}

Multiple patterns may have a common sub-pattern, which can be shared if they are searched in the same CUDA kernel.
On the other hand, mining multiple patterns simultaneously would need 
a significant amount of intermediate resources, e.g., registers, 
which results in low hardware utilization (occupancy) on GPU.

Instead of generating a single gigantic kernel for all patterns, 
we employ {\em kernel fission} to reduce register pressure.
Given multiple patterns, we leverage pattern analysis to find which patterns share the same sub-pattern, 
so that they should be merged into the same kernel to enjoy sharing. 
For those patterns do not share the same sub-patterns,
we generate different kernels for them,
so that each kernel is lightweight enough to avoid high register pressure.
For example, in 4-motifs (\cref{fig:motifs}), {\tt tailed-triangle}, 
{\tt diamond} and {\tt 4-clique} share the same sub-pattern {\tt triangle}.
So we generate a single CUDA kernel for the three patterns, 
in which they share the same workflow that enumerates triangles.
However, for the other patterns, 
since there is no sharing opportunity, we generate one kernel for each.
These separated kernels use fewer registers than a combined kernel,
so that each SM in GPU can accommodate more co-running warps to maximize utilization.
This improves performance by 15\% for mining 4-motifs.

\subsection{Support for Advanced Pruning Schemes} \label{subsect:pruning}

\begin{algorithm}[t]
\footnotesize
\caption{Pseudo code for counting {\tt diamond}}
\label{alg:diamond-count}
\begin{algorithmic}[1]
  \For{{\bf each} vertex $v_1 \in$ \vs in parallel}  \Comment{\hlc{match $v_1$ to $u_1$}}
    \For{{\bf each} vertex $v_2 \in$ \adj($v_1$)} \Comment{\hlc{match $v_2$ to $u_2$}}
	  \If{$v_2\ge v_1$} break; \Comment{\hlc{symmetry breaking}}
	  \EndIf
	  \State $n=$ |\adj($v_1$) $\cap$ \adj($v_2$)|; \Comment{\hlc{\# triangles incident to $(v_1, v_2)$}} \label{count-only:intersect}
      \State count += $n$*($n$-1)/2 \Comment{\hlc{choose 2 from $n$ to form a diamond}} \label{count-only:formula}
    \EndFor
  \EndFor
\end{algorithmic}
\end{algorithm}

\paragraph{(1) Counting-only Pruning.} \label{para:counting-only-pruning}
If the user is interested in \textit{counting} instead of \textit{listing} subgraphs,
there may exist an advanced pruning opportunity to further reduce the search space.
For example, to count edge-induced {\tt diamond}  (\cref{alg:diamond-count}), 
because a \texttt{diamond} consists of two triangles,
we first compute the triangle count $n$ for each edge $(v_1,v_2)$ 
using set intersection (Line \ref{count-only:intersect}),
and then use the formula 
$\binom{n}{2}$
$=n \times (n-1)/2$ 
to get the {\tt diamond} count (Line \ref{count-only:formula}). 
Note that this pruning opportunity is pattern specific and is not always available.
For example, there is no such opportunity for {\tt 4-cycle}.
Our pattern analyzer detects the opportunities by using automatic pattern decomposition~\cite{ESCAPE,DwarvesGraph}, and based on the detection, our code generator can accordingly generate the CUDA kernel.

\begin{figure}[t]
\centering
\includegraphics[width=0.48\textwidth]{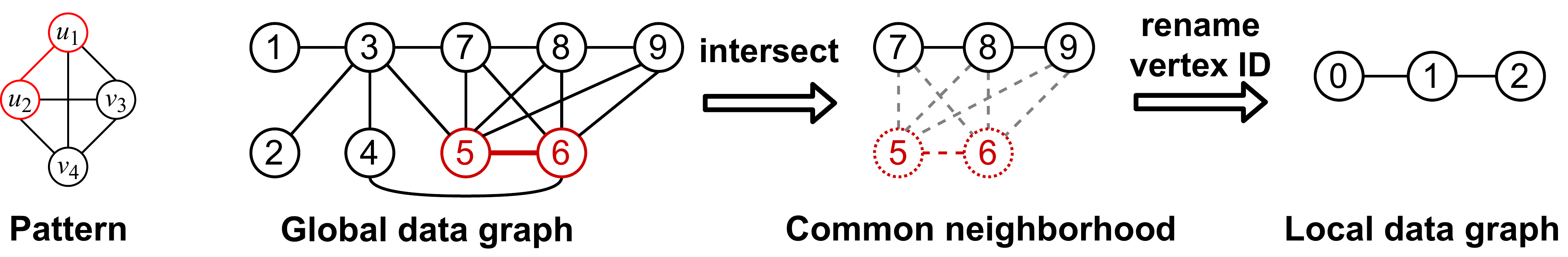}
\caption{Local graph constructed for $v_1$=5 and $v_2$=6 which are matched to hub vertices $u_1$ and $u_2$ in the pattern respectively. \hl{We first compute set intersection of vertex 5 and 6, to get their common neighbors (vertex 7, 8, 9). The common neighbors are renamed to form a local graph. Renaming can reduce \texttt{bitmap} storage. }}
\vspace{-9pt}
\label{fig:local-graph}
\end{figure}

\paragraph{(2) Local Graph Search (LGS).} \label{paragraph:lgs}
\hl{This is a pruning scheme used for \texttt{hub-patterns}. 
A hub-pattern contains at least one \texttt{hub} vertex that is connected to all other vertices.
For example, any vertex in a \texttt{clique} is a hub vertex.
The key idea of LGS is, instead of searching a massive data graph \dg, we can construct a small local graph for each vertex in \dg and search in the local graphs.
For a hub pattern with a hub vertex $u_1$, 
we match the first data vertex $v_1$ to $u_1$, 
and the entire sub-tree rooted by $v_1$ is confined within $v_1$'s 1-hop neighborhood. 
\cref{fig:local-graph} shows an example of constructing a local graph. 
Search in the local graph is faster because the vertex degrees in the local graph are smaller than those in the global data graph.}
When the pattern analyzer detects a hub-pattern,
the code generator inserts a call to construct local graphs, 
and generates code to search in the local graphs, instead of the original data graph.

\begin{itemize}

\item \hl{Previously, LGS has only been used for clique patterns~\cite{KClique}, while \sys generalizes and automates it for all hub patterns. 
Moreover, unlike CPUs, naive implementation on GPUs is not beneficial. 
We combine LGS with the \texttt{bitmap} format (see \cref{subsect:flexible-data}) to achieve significant speedups.}

\item 
{\it Input Awareness}. LGS is not always beneficial~\cite{Sandslash}. 
The key indicator is the maximum degree $\Delta$ of the data graph. 
For example, if $\Delta$ is too large, it is not beneficial due to high overhead of local graph construction.
Therefore, we generate CUDA kernels for both cases: LGS enabled and disabled. 
The runtime system checks if $\Delta$ is above a threshold and decides accordingly which kernel to use.
LGS brings us 1.2 $\sim$ 3.7$\times$ speedup on GPU for various data graphs.
\end{itemize}

\section{Device Primitives for Set Operations} \label{sect:primitives}

As \sys assigns each task to a warp, whenever there is a set operation,
all the thread in a warp work cooperatively to compute it.
For example, in \cref{alg:dfs-diamond}, there is a set intersection at \cref{algo:dfs:buffer}.
In \sys, set operations are done by invoking the corresponding device functions predefined in the GPU primitive library.
We leverage GPU hardware SIMD support to implement efficient set operations (\cref{subsect:simd}) and flexibly support various data formats for vertex sets (\cref{subsect:flexible-data}).

\subsection{SIMD-aware Primitives} \label{subsect:simd}

Given two sets $A$ and $B$, we need two major set operations in GPM:
(1) set intersection: $C = A\cap B$;
(2) set difference: $C=A-B$,
where $C$ is the output set.
Besides, another operation \textit{set bounding} is also often needed:
given a set $A$ and an upper bound $y$, set bounding computes $\{x | x < y \& x \in A$\}.
We discuss set intersection in detail, and the other operations are similar. 

In \cref{alg:dfs-diamond} \cref{algo:dfs:buffer},
the result of set intersection is stored in a buffer $W$ for reuse.
Buffering is widely used in GPM algorithms to avoid repetitive computation~\cite{AutoMine}.
To support buffering in \sys, each warp is allocated a private buffer in the GPU memory.
In the primitive functions, threads in a warp write outputs to the buffer in parallel.
To do this efficiently, we use CUDA warp-level primitives~\cite{warp-primitives} which are supported by the GPU hardware (special instructions).
For each vertex $v$ in set $A$, 
we use a boolean flag to indicate whether it exists in set $B$.
Using the flag, we compute a mask using {\tt \_\_ballot\_sync} primitive.
The mask is then used to compute the index and the total size of the buffer using {\tt \_\_popc} primitive. 

\noindent
\textbf{Implementation details.} Previous work has explored set intersection for SIMD~\cite{FESIA,QFilter,EvaluationSI,Manycore-Clique,FasterSI}
or GPU~\cite{Fox,TriCore,H-INDEX,list-intersect,si-gpu,batched-si,GPU-Merge,Merge,Merge-GPU,TC-GPU}.
We classify their algorithms into 3 categories: \textit{Merge-path}~\cite{GPU-Merge,Merge-GPU}, \textit{Binary-search}~\cite{Fox,TriCore} and \textit{Hash-indexing}~\cite{H-INDEX}.
We have extensively evaluated these methods on GPU, and we find that binary search works the best since it is less divergent.
In our library, we implement a high-performance binary search~\cite{TriCore}:
to exploit temporal locality, we leverage the scratchpad in GPU to pre-load the first five layers of the binary search tree, which further mitigates memory divergence.
We extend this method to also support set difference, set bounding, and local graph construction.

\subsection{Flexible Data Representation} \label{subsect:flexible-data}

\textit{Vertex set} is a key data structure in GPM, 
which is used for the neighbor list in \dg and the buffer $W$ in \cref{alg:dfs-diamond}.
Its representation on GPU has a major impact on performance.
For the set operations particularly, using a dense representation makes set operations easy to compute, but it requires more storage space.
If using a sparse representation, it saves space but complicates the computation of set operations.

We support two types of formats for vertex set on GPU: \texttt{sorted-list} (sparse) and \texttt{bitmap} (dense).
\texttt{sorted-list} is a list (i.e. array) of vertices sorted in ascending order. 
\texttt{bitmap} is a sequence of bits (length=|\vs|), each of which indicates the connectivity to a vertex in \vs. 
Set operations on \texttt{bitmap} are very simple and efficient,
but \texttt{bitmap} consumes more space when \vs is large.
Thus, by default we use \texttt{sorted-list}, and we only enable \texttt{bitmap} for hub-patterns since the \texttt{bitmap} size can be reduced significantly ($\Delta$ instead of |\vs|).

\section{Runtime Scheduling and Management} \label{sect:runtime}

Our runtime system is aware of the pattern, input data graph and GPU architecture to balance workload among multiple GPUs (\cref{sect:scheduler}) and make full use of the GPU memory (\cref{sect:mem-mgmt}).

\subsection{Task Scheduling for Multi-GPU} \label{sect:scheduler}

Given $n$ as the number of GPUs\footnote{We assume that every GPU has the same compute power for simplicity, otherwise it is not difficult to scale the workload by a factor accordingly.} available in the system and a data graph \dg with an edgelist
$\Omega = \{e_1, e_2, ..., e_m\}$ where $m=|$\es$|$ (in the case of symmetry breaking at level 2, $m=|$\es$|/2$),
the task scheduler aims to divide GPM {computation} onto the $n$ GPUs, 
by dividing $\Omega$ into $n$ segments, each of which has the same amount of work,
such that the execution time of the last completed GPU is minimized. 

BFS-based GPM systems, e.g., Arabesque, RStream, and Pangolin, balance workload by reassigning tasks at every level. But this does not work for the DFS algorithm because 
DFS does not work in the level-by-level way as BFS.
Existing DFS-based GPM systems target only CPUs, and thus can use sophisticated work stealing techniques~\cite{Fractal}. 
But this will incur non-trivial runtime overhead on multi-GPU ($\sim$20\%)~\cite{dynamicLB,TriCore}.

\begin{figure}[t]
\centering
\includegraphics[width=0.33\textwidth]{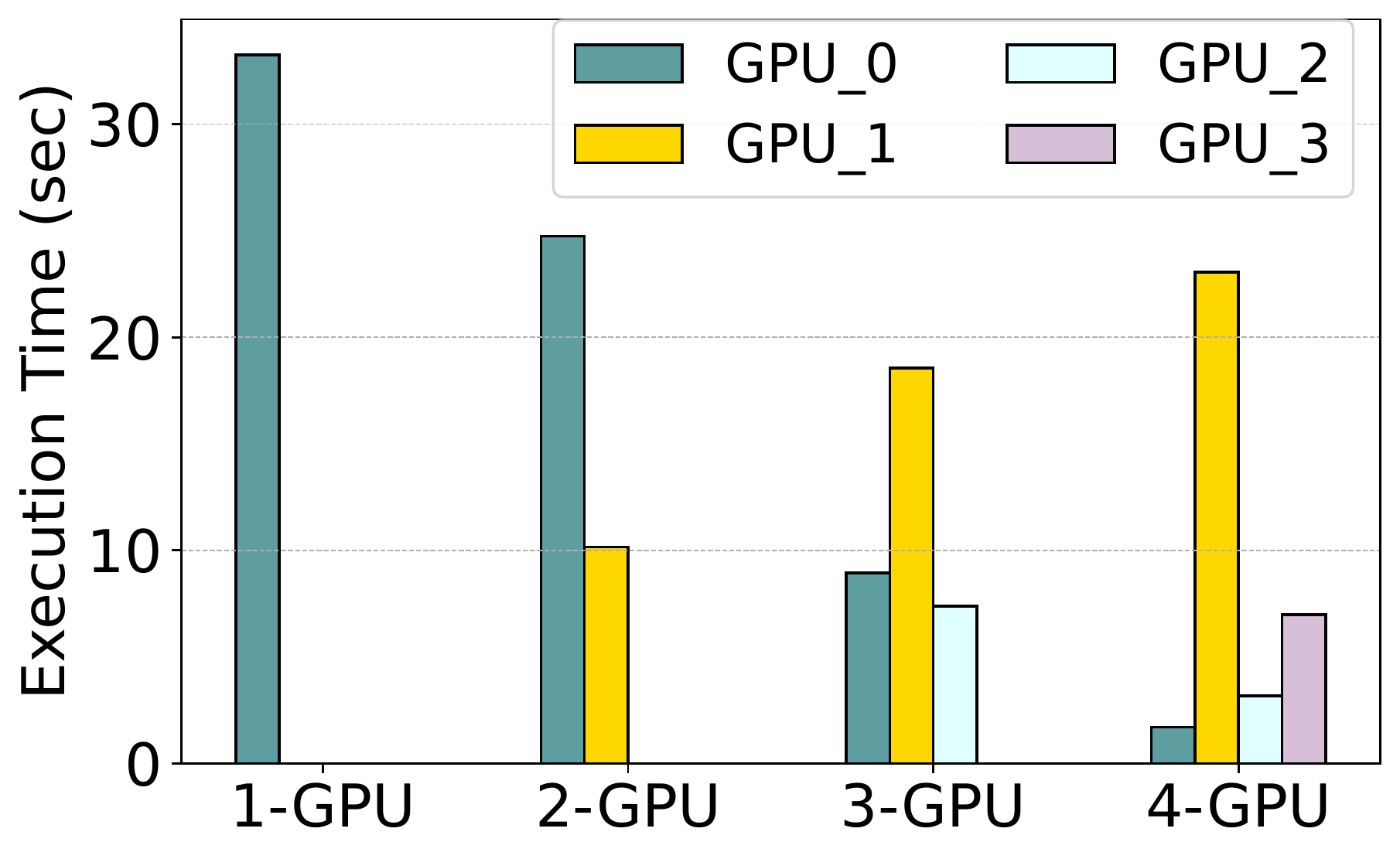}
\caption{Running time of each GPU using even-split: 3-MC on {\tt Tw2}.}
\vspace{-9pt}
\label{fig:even-split}
\end{figure}

\noindent
\textbf{Policy 1: Even-split Scheduling}.
$\Omega$ is to evenly split into $n$ consecutive ranges, each of which contains $m/n$ tasks.
This is used in existing triangle counting solvers on multi-GPU~\cite{H-INDEX}.
This policy is simple and has no scheduling overhead,
but it results in severe load imbalance for skewed graphs.
\cref{fig:even-split} shows the time spent on each GPU to finish its work under the even-split scheme.
Due to the skewness of the workload assigned to each GPU, 
under the 2-GPU setting we observe that GPU\_0 takes much more time to finish its work than GPU\_1.
The same time variance is observed for the 3-GPU and 4-GPU setting.
Worse still, in the 4-GPU setting, since most of the heavy tasks are assigned to GPU\_1, it makes the 4-GPU setting even slower than the 3-GPU setting.
This means the even-split scheme does not scale beyond 3-GPU for this benchmark.
The reason of poor scalability is two-folds: (1) the granularity of splitting workload is too coarse-grain; (2) it is unaware to the skewness of task workload by assuming every task has the same amount of work. 

\noindent
\textbf{Policy 2: Round-robin Scheduling}. Each GPU has a task queue, denoted as $Q_i$ for the $i$-th GPU, $i\in[0,n)$. 
The tasks in $\Omega$ are assigned to each queue in a round-robin fashion, i.e., $e_j$ is assigned to $Q_{i}$, where $i=j$ {\tt mod} $n$, $j\in[0,m)$.
This is a fine-grained scheduling policy that has been used in existing motif counting solvers on multi-GPU~\cite{Rossi}.
The policy comes with some overhead, i.e., copying tasks into task queues.
This copy is needed only once for a specific data graph and $n$, i.e., once done, the queues can be reused for mining different patterns.

\noindent
\textbf{Policy 3: Chunked Round-robin Scheduling}.
$\Omega$ is first split into lots of small chunks, and then we assign chunks to the task queues in a round-robin way. 
This is a generalization of the previous two policies. When the chunk size $c=m/n$, it becomes the same as policy 1.
When $c=1$, it becomes the same as policy 2.
Thus if $c$ is too small the data copying overhead will be high, but if $c$ is too large, we see load imbalance as in policy 1. 
We use $c=\alpha\times y$, where $y$ is the total number of warps and $\alpha$ is a constant (set to 2 empirically). 
Our chunking is also pattern aware, as described in \cref{sect:mem-mgmt}.
\textit{Implementation details}. 
To further reduce data copy overhead, we parallelize it as the location to copy to is fixed for each queue if the chunk size is fixed. 
Note that this overhead is constant to the pattern size $k$, 
which is trivial (< 1\%) when $k>3$, since the GPM computation is exponential to $k$.
For small pattern like triangle,
we overlap the scheduling overhead with the GPU computation, by first assigning a few chunks to each GPU and launch the kernel.
During the GPU computation, we continue sending the remaining chunks from the CPU to feed the GPUs.
Orthogonal work on ordering tasks in $\Omega$~\cite{Rossi} or grouping tasks by community may help further improve load balance and locality. 

\subsection{GPU Memory Management} \label{sect:mem-mgmt}
GPU memory is a scarce resource. 
In GPM algorithms, the major memory usage involves the data graph \dg, the edgelist $\Omega$ and the buffers  (e.g., $W$ in \cref{alg:dfs-diamond})). 
For FSM, the subgraph list of each pattern requires additional space.


\noindent
\textbf{(1) Preprocessing the data graph}. 
We have discussed \textit{orientation} in \cref{subsect:interface}.
In the multi-GPU setting, for any hub-pattern, since the search is confined in the root vertex $v_1$'s neighborhood, we partition \vs into $n$ subsets ($n$ is the number of GPUs).
For each $i$-th subset we generate its vertex induced subgraph of \dg,
and copy it to the $i$-th GPU.
This partitioning reduces memory usage and guarantees that there is no communication needed between GPUs.
This technique has been used in ~\cite{DistTC} only for triangle counting. 
We generalize it for all hub-pattern problems.
The scheduling policy is then adjusted by chunking vertices and assigning incident edges in $\Omega$ to the corresponding GPUs.
For non-hub patterns, we do not partition \dg if it can fit in the single-GPU memory. 
This is because GPM algorithms access multi-hop neighbors, 
which leads to non-trivial communication overhead~\cite{DistGraph},
especially for small-diameter graphs.
When \dg is too large to fit in memory, we leverage community-aware partition~\cite{METIS} to minimize communication.

\noindent
\textbf{(2) Reducing the size of edgelist}.
For $\Omega$, we apply an important optimization by considering symmetry at the edge level (level-2).
Since \dg is an undirected graph, for each undirected edge in \dg, 
the edgelist contains two instances, 
each for one of the two directions of the edge. 
However, when there is a partial order between $v_1$ and $v_2$ for symmetry breaking, 
we generate the edgelist that contains only one instance. 
More specifically, if $v_1 > v_2$ is included in the symmetry order (e.g., in \cref{fig:symmetry} (b)),
the edgelist includes only the edges whose source vertex id is larger than its destination vertex id.
In this way, we can reduce half of the edges before execution.
It not only saves memory but also reduces checking on-the-fly.
Note that there is a similar optimization~\cite{LIGHT} to split the neighbor list of each vertex $v$ into two sets, with one holding all neighbors whose IDs are larger than $v$, and the other holding the rest which have smaller IDs than $v$.
This reduces on-the-fly checking, but it is not used to reduce memory usage. 

\noindent
\textbf{(3) Adaptive buffering}. 
In \sys's warp-centric DFS walk, each warp is allocated with $X$ buffers.
The value of $X$ is pattern specific and
the pattern analyzer can decide it when generating the search plan. 
For a pattern of size $k$, $X \leq k-3$ because the first two levels and the last level do not need buffers. 
So the worst case memory consumption for buffering is $O(\Delta \times (k-3))$.
This is linear to $k$ for a given specific data graph.
In comparison, the intermediate data generated in Pangolin is exponential to $k$, which can be easily over the GPU memory capacity (see in \cref{sect:compare-gpu}).
Although $\Delta$ is much smaller than \es (see \cref{tab:input}), given the large number of warps in GPU, 
the memory space for buffers can still be very large.
Therefore, the runtime limits the total number of warps to save memory usage, 
so that all tasks assigned to the same warp share the buffer usage.
In this way, given different data graphs, we can adaptively tune the number of warps to make full use of memory and maximize parallelism.
More specifically, we subtract the size of \dg and $\Omega$ from the total GPU memory size, to get the remaining memory size, denoted as $Y$. 
Then we can get the maximum number of warps $Y/(X\times\Delta)$.
Finally we launch \texttt{min}($Y/(X\times\Delta)$, |$\Omega$|) warps.

\noindent
\textbf{(4) Reducing memory allocation using label frequency}. 
This optimization is particularly useful for problems that find \textit{frequent patterns}, such as FSM.
The graph loader in \sys computes the vertex frequency for each label. 
This information can be leveraged to find \textit{frequent labels}, i.e., labels with vertex frequency above the user-defined support threshold $\sigma_{min}$.
Since infrequent labels can not be part of frequent patterns, the total number of possible {frequent patterns} $N$ can be significantly reduced, if there are many infrequent labels.
Note that in FSM we allocate a \textit{subgraph list} for each possible pattern to store subgraphs for aggregation, 
and the memory consumption of these subgraph lists is proportional to $N$.
With this awareness of the input (i.e., label frequency), we can drastically reduce this memory consumption in many cases.


\begin{table}[t]
\centering
\resizebox{0.42\textwidth}{!}{
	\begin{tabular}{crrrrrr}
		\Xhline{2\arrayrulewidth}
		\bf{Graph} & \bf{Source} & {\bf{|V|}} & {\bf{|E|}} & {\bf{Label}} & {\bf{Max deg.} $\Delta$}\\
		\hline
		\texttt{Mi} & {Mico}~\cite{GraMi} & 0.1M & 2M & 29 & 1,359 \\
		\texttt{Pa} & {Patents}~\cite{Patent} & 3M & 28M & 37 & 789 \\
		\texttt{Yo} & {Youtube}~\cite{Youtube} & 7M & 114M & 28 & 4,017 \\
		\texttt{Lj} & {LiveJournal}~\cite{SNAP} & 4.8M & 43M & 0 & 20,333\\
		\texttt{Or} & {Orkut}~\cite{SNAP} & 3.1M & 117M & 0 & 33,313\\
		\texttt{Tw2} & {Twitter20}~\cite{Konect} & 21M & 530M & 0 & 698,112\\
		\texttt{Tw4} & {Twitter40}~\cite{twitter40} & 42M & 2,405M & 0 & 2,997,487\\
		\texttt{Fr} & {Friendster}~\cite{friendster} & 66M & 3,612M & 0 & 5,214\\
		\texttt{Uk} & {Uk2007}~\cite{uk2007} & 106M & 6,603M & 0 & 975,419\\
		\Xhline{2\arrayrulewidth}
	\end{tabular}
}
\caption{\small Data graphs (symmetric, no loops or duplicate edges). Maximum degrees are smaller when orientation is applied for cliques.}
\label{tab:input}
\end{table}

\section{Evaluation}\label{sect:eval}


We compare \sys 
\footnote{\sys source code: https://github.com/chenxuhao/GraphMiner} 
with state-of-the-art systems: 
(1) GPM system on GPU, Pangolin~\cite{Pangolin}, (2) subgraph matching solver on GPU, \pbe \cite{GPU-Subgraph,reuse-subg},
(3) CPU-based GPM system Peregrine~\cite{Peregrine} and 
(4) CPU-based subgraph matching system GraphZero~\cite{GraphZero,AutoMine}.
Note that Pangolin also provides a CPU implementation, but it is slower than GraphZero.

\cref{tab:input} lists the data graphs.
The first 3 graphs ({\tt Mi}, {\tt Pa}, {\tt Yo}) are vertex-labeled graphs which are used for FSM.
We use all the GPM applications listed in \cref{subsect:define} for evaluation, i.e., TC, $k$-CL, SL, $k$-MC. 
For SL, we use two patterns {\tt 4-cycle} and {\tt diamond}. 
Note that GraphZero does not support FSM, 
Pangolin does not support SL, and \pbe does not support $k$-MC and FSM.
For FSM, we include DistGraph~\cite{DistGraph} in \cref{tab:fsm} as the state-of-the-art hand-written FSM solver.

CPU-based systems and solvers are evaluated on a 4 socket machine with Intel Xeon Gold 5120 2.2GHz CPUs (56 cores in total) and 190GB RAM,
while GPU-based solutions are evaluated on NVIDIA V100 GPUs (each with 32GB device memory).
We exclude preprocessing (e.g., DAG construction in Pangolin
and vertex reordering in Peregrine) time in all systems.
We use a time-out of 30 hours for CPU and 8 hours for GPU,
and report all results as an average of three runs.
We show single-GPU performance in \cref{sect:compare-gpu} and compare with CPU solutions in \cref{sect:compare-cpu}.
Multi-GPU performance of \sys is shown in \cref{sect:multi-gpu}.
Impact of optimizations is analyzed in \cref{subsect:gpu-util}.

\begin{table}[t]
\footnotesize
\centering
\resizebox{0.42\textwidth}{!}{
\begin{tabular}{r|rrrrrr}
\Xhline{2\arrayrulewidth}
\textbf{Data Graph} & \textbf{Lj} & \textbf{Or} & \textbf{Tw2} & \textbf{Tw4} & \textbf{Fr} & \textbf{Uk} \\ \hline
\sys (GPU)      & 0.03 & 0.14 & 1.6  & 5.1 & 3.2 & 7.5 \\
Pangolin (GPU)  & 0.06 & 0.25 & 3.0  & OoM & 5.2 & OoM \\
PBE (GPU)   & 0.27 & 1.12 & 13.4 & 53.5 & 23.0 & 55.3\\
Peregrine (CPU) & 1.63 & 7.25 & 112.1& 8492.4 & 100.3 & 3640.9 \\
GraphZero (CPU) & 0.61 & 2.22 & 24.4 & 1399.3 & 49.0 & 1041.3 \\
\Xhline{2\arrayrulewidth}
\end{tabular}
}
\caption{TC running time (sec). OoM: out of memory.}
\vspace{-13pt}
\label{tab:tc}
\end{table}

\begin{table}[t]
\centering
\resizebox{0.45\textwidth}{!}{
\begin{tabular}{c|rrrrr|rrr}
\Xhline{2\arrayrulewidth}
\textbf{Pattern} & \multicolumn{5}{c|}{\textbf{4-CL}} & \multicolumn{3}{c}{\textbf{5-CL}} \\ \hline
\textbf{Data Graph} & \textbf{Lj} & \textbf{Or} & \textbf{Tw2} & \textbf{Tw4} & \textbf{Fr} & \textbf{Lj} & \textbf{Or} & \textbf{Fr} \\ \hline
\sys (G) & 0.32 & 0.54 & 113.3 & 362.9 & 7.3 & 3.2 & 1.7 & 13.1 \\
Pangolin (G)  & 1.48 & 4.04 & OoM & OoM & OoM & OoM & OoM & OoM \\
PBE (G)   & 3.90 & 11.11 & 3640.1 & TO & 117.8 & 246.4 & 99.2 & 399.8 \\
Peregrine (C) & 15.90 & 73.70 & 39921.0 & TO & 397.3 & 520.8 & 782.1 & 957.6 \\
GraphZero (C) & 3.48 & 12.96 & 2152.2 & 20591.1 & 177.7 & 60.0 & 48.3 & 243.3 \\
\Xhline{2\arrayrulewidth}
\end{tabular}
}
\caption{\small $k$-CL running time (sec). TO: timed out.} 
\label{tab:cl}
\end{table}

\begin{table}[t]
\resizebox{0.48\textwidth}{!}{
\begin{tabular}{c|rrrrr|rrr}
\Xhline{2\arrayrulewidth}
\textbf{Pattern}    & \multicolumn{5}{c|}{\textbf{Diamond}} & \multicolumn{3}{c}{\textbf{4-cycle}}    \\ \hline
\textbf{Data Graph} & \textbf{Lj} & \textbf{Or} & \textbf{Tw2}         & \textbf{Tw4} & \textbf{Fr} & \textbf{Lj} & \textbf{Or} & \textbf{Fr} \\ \hline
\sys (G) & 0.29 & 0.75 & 26.8 & 183.1 & 12.8 & 2.7 & 33.7 & 1291.2 \\
PBE (G)  & 0.48 & 1.71 & 26.3 & 102.0 & 39.9 & 17.3 & 177.8 & 5211.3 \\
Peregrine (C) & 5.38 & 10.24 & 553.6 & 20898.4 & 178.1 & 144.4 & 1867.2 & 32276.8 \\
GraphZero (C) & 1.73 & 7.27 & 165.1 & 7938.6 & 136.4 & 34.0 & 345.5  & 9251.5 \\
\Xhline{2\arrayrulewidth}
\end{tabular}
}
\caption{\small SL running time (sec). `G': GPU; `C': CPU.} 
\label{tab:sgl}
\end{table}

\begin{table}[t]
\resizebox{0.49\textwidth}{!}{
\begin{tabular}{r|rrrrr|rrr}
\Xhline{2\arrayrulewidth}
\multicolumn{1}{c|}{\textbf{Pattern}} & \multicolumn{5}{c|}{\textbf{3-Motif}} & \multicolumn{3}{c}{\textbf{4-Motif}}    \\ \hline
\textbf{Data Graph \ } & \textbf{Lj} & \textbf{Or} & \textbf{Tw2} & \textbf{Tw4} & \textbf{Fr} & \textbf{Lj} & \textbf{Or} & \textbf{Fr} \\ \hline
\sys (G) & 0.17 & 0.97 & 33.3 & 1703.6 & 22.0 & 138.1 & 2068.4 & 15475.4 \\
Pangolin (G) & 2.05 & 22.62 & 1165.5 & OoM & OoM & OoM & OoM & OoM \\
Peregrine (C) & 9.36 & 19.46 & 418.7 & 27954.9 & 367.9 & 1435.4 & 20219.1 & TO \\
GraphZero (C) & 1.50 & 7.74 & 276.5 & 7439.4 & 169.6 & 3039.6 & 16394.6 & TO \\
\Xhline{2\arrayrulewidth}
\end{tabular}
}
\caption{\small $k$-MC running time (s). OoM: out of mem.; TO: timed out.}
\label{tab:mc}
\end{table}

\subsection{Single-GPU Performance} \label{sect:compare-gpu}
We compare with Pangolin and PBE on a V100 GPU.
\cref{tab:tc} lists the GPU running time for triangle counting (TC).
We observe that Pangolin runs out of memory for {\tt Tw4}\footnote{Since data graphs are oriented in TC, {\tt Fr} takes less memory than {\tt Tw4}} and {\tt Uk},
while \sys can run with all the data graphs.
We also observe that \sys is constantly faster than Pangolin, 
due to optimized set operations in our library.
On average, \sys is 1.8$\times$ faster than Pangolin on V100 GPU.

The speedups are more significant for $k$-CL and $k$-MC.
As shown in \cref{tab:cl}, \sys outperforms Pangolin by 4.6$\times$ and 7.6$\times$ for 4-clique listing on {\tt Lj} and {\tt Or} respectively.
The speedups mainly come from data reuse enabled in DFS (i.e., buffering $W$ in \cref{alg:dfs-diamond}) and optimized set operations\footnote{It can not be directly used for Pangolin, as Pangolin maps \textit{connectivity checks}~\cite{Pangolin} to threads, but \sys maps set operations to warps.}.
Meanwhile, for all the rest of graphs and the larger pattern 5-clique,
Pangolin runs out of memory.
Similar trend is found in \cref{tab:mc}, where we observe an average of 21.3$\times$ speedup over Pangolin on 3-MC,
and Pangolin also runs out of memory for most of the cases.
\sys managed to run all cases, which demonstrates that its DFS order and optimization J and K in \cref{tab:optimizations} can effectively reduce memory consumption.

For FSM in \cref{tab:fsm}, \sys is competitive with Pangolin for the small graphs, since we use bounded BFS (optimization M in \cref{tab:optimizations}) that provides enough parallelism. 
For the largest graph {\tt Yo}, Pangolin runs out of memory again, while \sys succeeds to run it, thanks to both
optimization M and N in \cref{tab:optimizations} which help reduce memory consumption. 

Overall, 
\sys achieves an average speedup of \textbf{5.4}$\times$ over Pangolin, 
and the speedup is more significant for larger patterns.
Moreover, \sys can run much larger graphs.

\begin{figure*}[t]
\centering
\begin{subfigure}[b]{0.3\textwidth}
\includegraphics[width=0.95\textwidth]{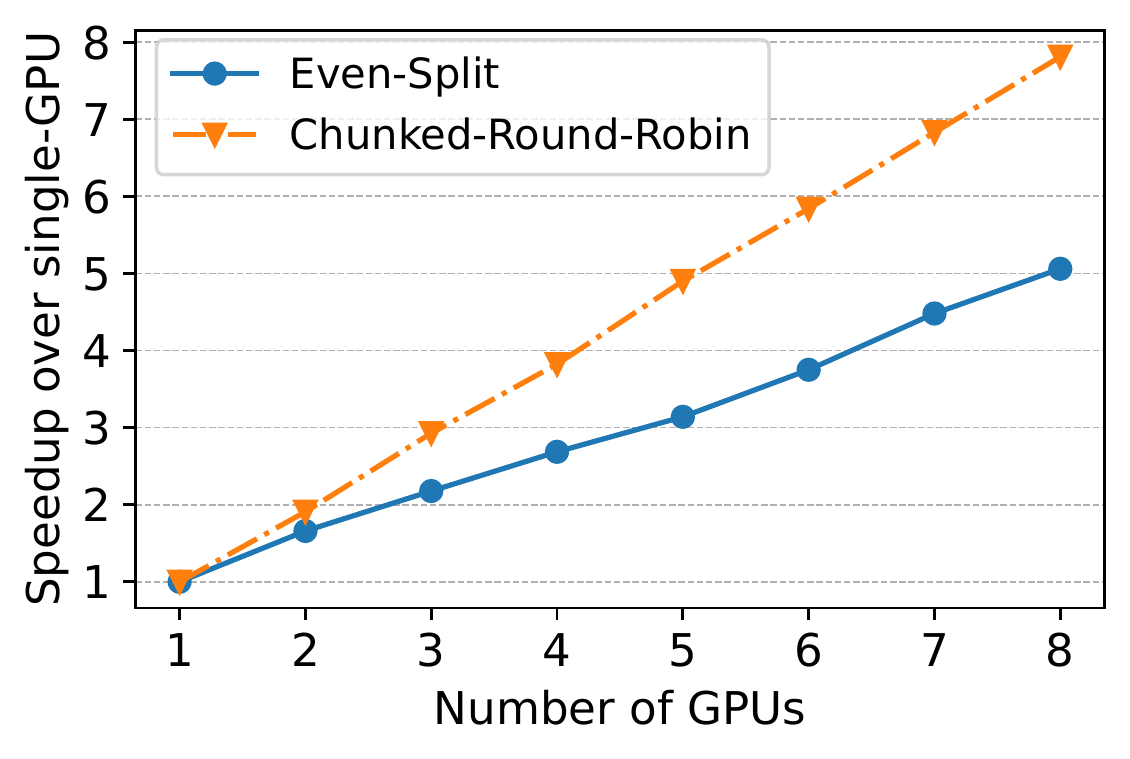}
\vspace{-9pt}
\caption{Triangle counting on {\tt Tw4}.}
\label{fig:multigpu1}
\end{subfigure}
\hfill
\begin{subfigure}[b]{0.3\textwidth}
\includegraphics[width=0.95\textwidth]{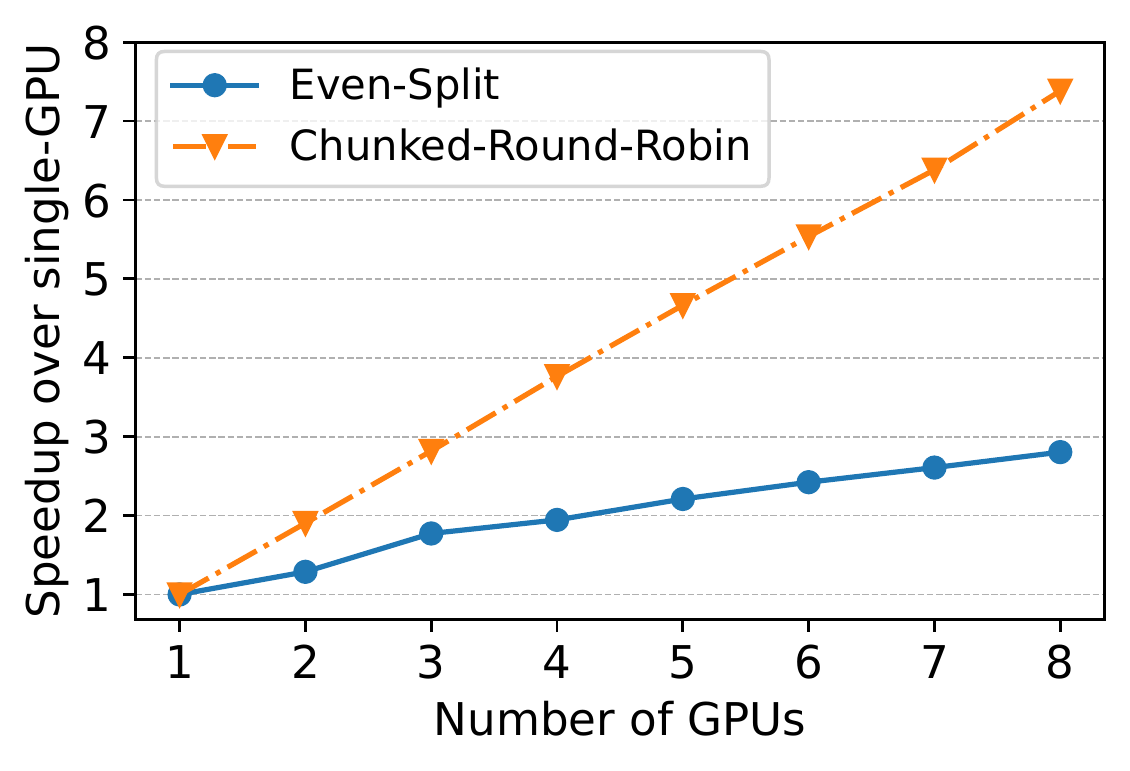}
\vspace{-9pt}
\caption{Listing {\tt 4-cycle} on {\tt Fr}.}
\label{fig:multigpu2}
\end{subfigure}
\hfill
\begin{subfigure}[b]{0.3\textwidth}
\includegraphics[width=0.95\textwidth]{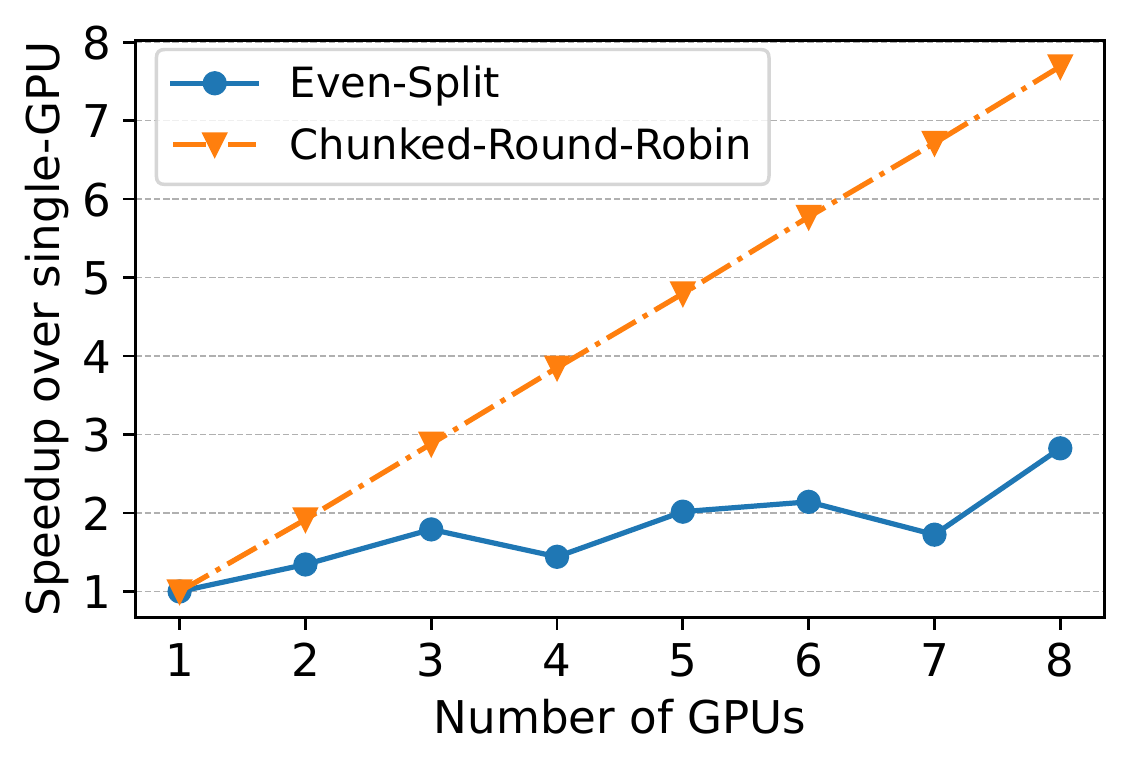}
\vspace{-9pt}
\caption{3-motif counting on {\tt Tw2}.}
\label{fig:multigpu3}
\end{subfigure}
\caption{\sys multi-GPU scalability using two task scheduling policies: even-split vs. chunked-split.}
\label{fig:three graphs}
\end{figure*}

We also compare with \pbe~\cite{GPU-Subgraph,reuse-subg} on the V100 GPU.
\pbe partitions the data graph when it gets large, which allows it run all the single-pattern workloads.
However, its performance is even worse (3.8$\times$ slower) than Pangolin, 
due to the cross-partition communication overhead and lack of data graph orientation.
Particularly, for subgraph listing, as \texttt{diamond} contains a sub-pattern \texttt{triangle} but \texttt{4-cycle} does not, searching \texttt{diamond} generates much less intermediate data than searching \texttt{4-cycle}.
Thus in \cref{tab:sgl} we observe that PBE's \texttt{4-cycle} performance is much worse than \sys as it has to do partitioning and suffers from the overhead.
Overall, \sys achieves a \textbf{7.2}$\times$ speedup over \pbe on average.

\subsection{Mining on GPU vs. on CPU} \label{sect:compare-cpu}

To evaluate how much speedup we can get from GPU over CPU,
we compare \sys (on V100 GPU) with GraphZero (on 56-core CPU).
Note that for each specific GPM application, \sys and GraphZero use exactly the same matching order and symmetry order, 
making it a fair comparison to show the benefit from the difference of hardware architectures.
As listed in \cref{tab:tc}, \sys is significantly faster than GraphZero on TC, with an average speedup of 38.0$\times$. 
The same trend is observed for $k$-CL in \cref{tab:cl},
where \sys outperforms GraphZero by 18.2$\times$.
This tremendous performance improvement is due to three parts:
(1) the orientation optimization, 
(2) higher throughput (i.e. more parallelism) on GPU,
and (3) our high-performance set operations on GPU.

\begin{figure*}
\centering
\begin{minipage}[]{0.27\textwidth}
\includegraphics[width=\textwidth]{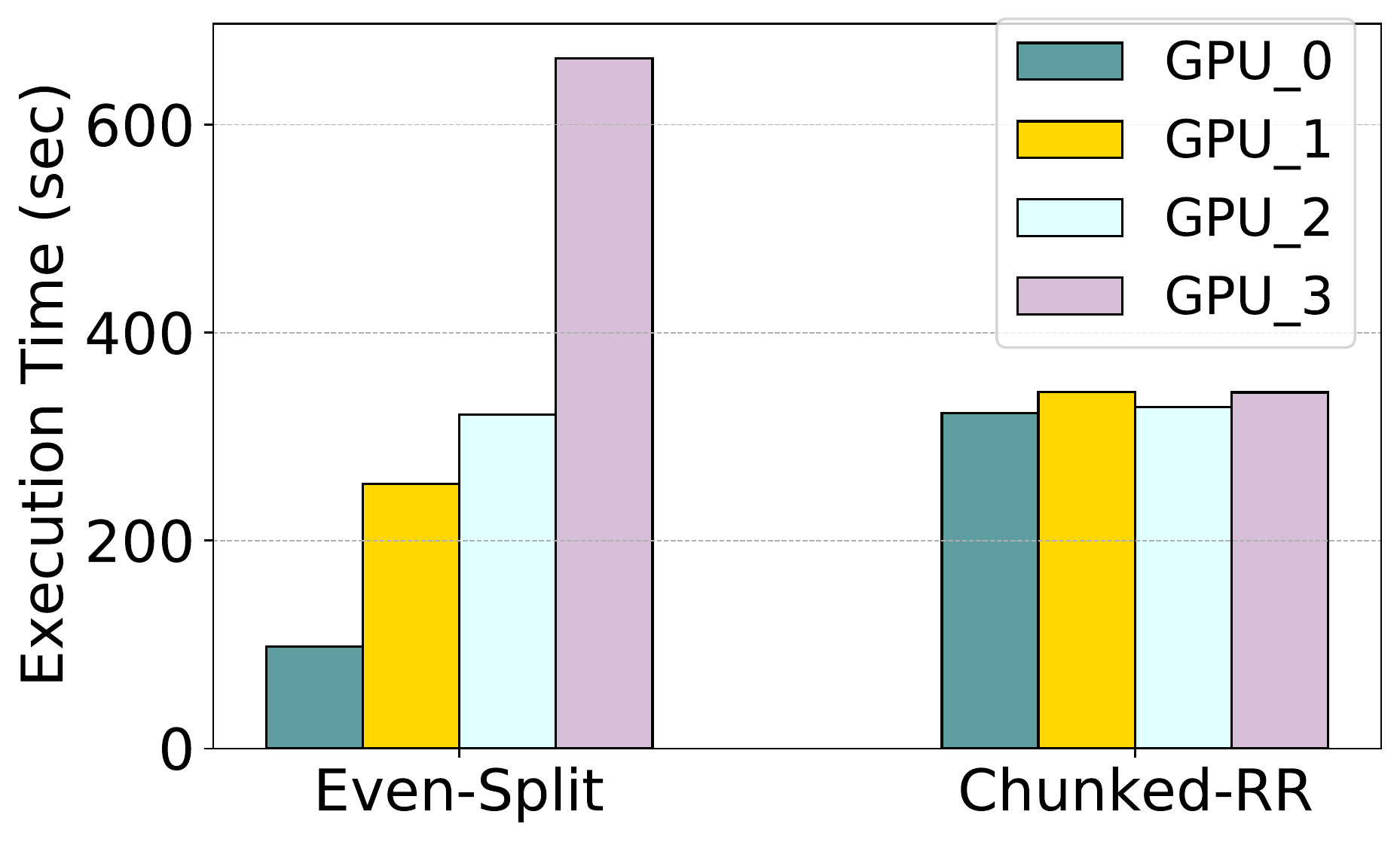}
\caption{Running time of each GPU in the 4-GPU setting: {\tt 4-cycle} on {\tt Fr}.}
\label{fig:distr}
\end{minipage}
\hfill
\begin{minipage}[]{0.27\textwidth}
\includegraphics[width=\textwidth]{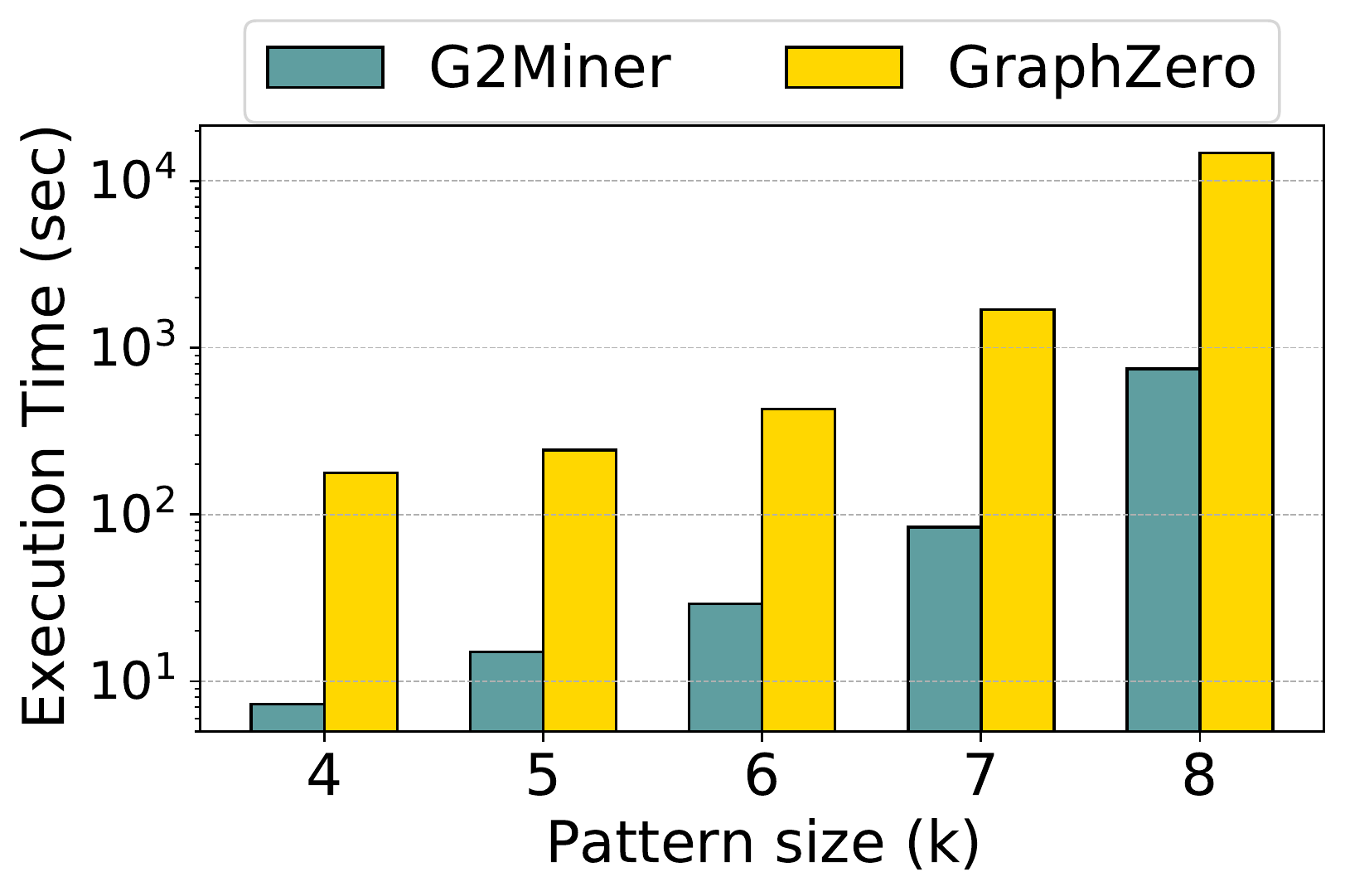}
\vspace{-18pt}
\caption{Running time of $k$-clique listing over {\tt Fr}, $k\in$[4,8].}
\label{fig:k-clique}
\end{minipage}
\hfill
\begin{minipage}[]{0.39\textwidth}
\includegraphics[width=\textwidth]{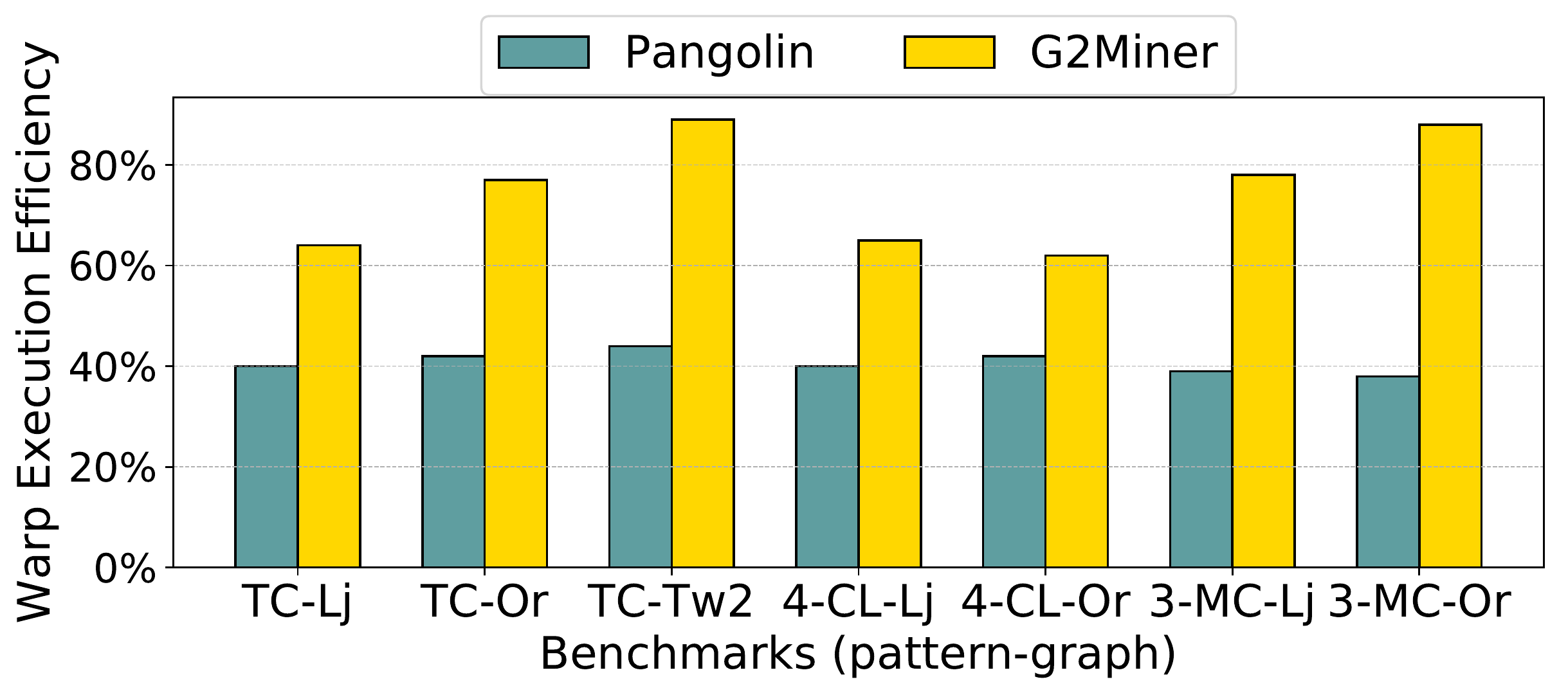}
\caption{Warp execution efficiency.}
\label{fig:warp-eff}
\end{minipage}
\end{figure*}

For SL, orientation can not be applied. 
Thus it can be used to evaluate the benefit of the other two parts.
As shown in \cref{tab:sgl}, \sys still achieves overwhelmingly better performance than GraphZero, with an average speedup of 10.5$\times$.
The speedup would be marginal if we use the BFS strategy in Pangolin and PBE or implement our DFS scheme naively.

While TC, $k$-CL and SL uses only set intersection, 
$k$-MC includes both set intersection and set difference.
As \sys optimizes both operations, 
we also observe dramatic performance boost for $k$-MC. 
In \cref{tab:mc}, it constantly outperforms GraphZero for all benchmarks.
On average \sys is 8.5$\times$ faster than GraphZero.

Overall, \sys on GPU achieves \textbf{15.2}$\times$ speedup over GraphZero on CPU, 
which demonstrates the significant benefit of using GPU to accelerate GPM applications.

As GraphZero does not support FSM, we also compared to Peregrine. 
\sys on GPU is \textbf{48.3}$\times$ faster than Peregrine on CPU.
Note that Peregrine does not mine multiple patterns simultaneously for multi-pattern problems.
Instead, for $k$-MC and FSM, it enumerates every pattern one by one, making it impossible to reuse data across similar patterns.
Thus it is mostly even slower than GraphZero.

\begin{table}[t]
\resizebox{0.49\textwidth}{!}{
\begin{tabular}{c|rrrr|rrrr|rrrr}
\hline
\multicolumn{1}{l|}{\textbf{Data Graph}}  & \multicolumn{4}{c|}{\textbf{Mico}}                          & \multicolumn{4}{c|}{\textbf{Patent}}                        & \multicolumn{4}{c}{\textbf{Youtube}}                        \\ \hline
\textbf{$\sigma$} & \textbf{300} & \textbf{500} & \textbf{1000} & \textbf{5000} & \textbf{300} & \textbf{500} & \textbf{1000} & \textbf{5000} & \textbf{300} & \textbf{500} & \textbf{1000} & \textbf{5000} \\ \hline
\textbf{G2Miner (G)}   & 0.6          & 0.4          & 0.3           & 0.1           & 2.6          & 2.6          & 2.6           & 1.7           & 7.2          & 6.0          & 6.0           & 8.7           \\
\textbf{Pangolin (G)}  & 0.6          & 0.5          & 0.3           & 0.2           & 2.7          & 2.7          & 2.7           & 1.7           & OoM          & OoM          & OoM           & OoM           \\
\textbf{Peregrine (C)} & 4.4          & 4.4          & 4.2           & 4.3           & 94.2         & 103.8        & 118.4         & 94.3          & 59.3         & 52.8         & 69.9          & 60.8          \\
\textbf{DistGraph (C)} & 56.1 & 61.0 & 57.6 & 57.0 & 13.2 & 13.1 & 13.0 & 14.1 & OoM  & OoM & OoM  & OoM \\ 
\hline
\end{tabular}
}
\caption{3-FSM running time (sec). OoM: out of memory. }
\label{tab:fsm}
\end{table}

\subsection{Multi-GPU Scalability} \label{sect:multi-gpu}

We evaluate multi-GPU performance by varying the number of GPUs from 1 to 8 in a single machine.
Since \pbe and Pangolin do not support multi-GPU, 
we only evaluate \sys in this section.
We compare two task scheduling policies in \cref{fig:three graphs}.
As illustrated, the chunked round-robin scheme constantly works much better than the even-split scheme.
More importantly, the chunked scheme scales linearly for all cases,
while the even-split scheme fails to scale beyond 3-GPU for 3-MC on {\tt Tw2}.
The poor scalability of even-split is dues to the load imbalance.
As shown in \cref{fig:distr}, in the 4-GPU setting, the execution time of each GPU varies dramatically for the even-split setting.
In contrast, for the chunked scheme, each GPU finishes its work roughly at the same time.

\subsection{Impact of Optimizations} \label{subsect:gpu-util}

\begin{table*}[t]
\footnotesize
\centering
\begin{tabular}{c|rrrrr|rrrrr|rrr}
\Xhline{2\arrayrulewidth}
\textbf{Pattern} & \multicolumn{5}{c|}{\textbf{Diamond}} & \multicolumn{5}{c|}{\textbf{3-Motif}} & \multicolumn{3}{c}{\textbf{4-Motif}} \\ \hline
\textbf{Time (sec)} & \textbf{Lj} & \textbf{Or} & \textbf{Tw2} & \textbf{Tw4} & \textbf{Fr} & \textbf{Lj} & \textbf{Or} & \textbf{Tw2} & \textbf{Tw4} & \textbf{Fr} & \textbf{Lj} & \textbf{Or} & \textbf{Fr} \\ \hline
\sys (GPU) & 0.09 & 0.47 & 9.9 & 66.9 & 10.4 & 0.06 & 0.27 & 6.8 & 21.4 & 5.2 & 2.6 & 34.2 & 1307.2 \\
Peregrine (CPU) & 2.20 & 8.66 & 245.8 & 16312.6 & 158.8 & 2.51 & 4.90 & 116.0 & 8447.4 & 165.3 & 163.6 & 1701.4 & TO \\
\Xhline{2\arrayrulewidth}
\end{tabular}
\caption{\hl{Running time of \sys vs. Peregrine, both with counting-only pruning enabled. TO: timed out.}}
\vspace{-13pt}
\label{tab:count}
\end{table*}

\hl{Different optimizations in \cref{tab:optimizations} contribute differently to the performance improvement. 
First, architecture-aware optimizations are crucial for all workloads on GPU. 
\sys is 5.4$\times$ faster then Pangolin, where \textit{two-level parallelism} (C in \cref{tab:optimizations}) and \textit{SIMD-aware primitives} (H in \cref{tab:optimizations}) contribute 3.1$\times$ and 1.7$\times$ respectively.
Second, for a pattern-aware optimization, it is beneficial only for the target pattern(s), 
and the speedups vary a lot depending on how much the search space is pruned. 
For example, \textit{local-graph search} (E+F in \cref{tab:optimizations}) brings 1.2$\times \sim$ 3.7$\times$ speedup for \texttt{hub-patterns} (2.1$\times$ on average),
while \textit{counting-only pruning} (D in \cref{tab:optimizations}) achieves 1.2$\times$ (\texttt{diamond}, \texttt{Fr}) to 79.7$\times$ (\texttt{3-motif}, \texttt{Tw40}), with 6.2$\times$ on average.
Other optimizations in \cref{tab:optimizations} are for memory saving, which is crucial for enabling larger datasets.}

\noindent
{\bf Large Pattern and Large Graph}.
A major advantage of \sys over Pangolin is that \sys can support much larger graphs and patterns. 
\cref{fig:k-clique} shows that \sys can run up to 8-clique listing on a billion-edge graph {\tt Fr}.
In contrast, Pangolin can not even run 4-clique due to out-of-memory, as shown in \cref{tab:cl}.
\cref{fig:k-clique} also shows that, from 4-clique to 8-clique, \sys on GPU consistently achieves an order of magnitude speedup over GraphZero on the CPU, 
although the GPU has much less memory than the CPU.
This trend implies that GPUs can be not only capable but also highly efficient for processing large graphs and patterns, 
thanks to \sys{'s} memory management and optimizations for the GPU architecture.

\noindent
{\bf GPU Efficiency}.
To evaluate GPU utilization, we measure {\it warp execution efficiency},
which is the average percentage of active threads in each executed warp.
As shown in \cref{fig:warp-eff}, the warp execution efficiency in Pangolin is around 40\%. 
This is relatively low since more than half of the compute horse power is wasted.
In comparison, \sys significantly improves the warp execution efficiency. 
This is mainly due to the highly efficient implementation of our warp-centric set operations.
Besides, we also measure \textit{branch efficiency}, i.e.,
the ratio of non-divergent branches to total branches. 
Although \sys uses DFS, We find that Pangolin and \sys have almost the same branch efficiency, thanks to the two-level parallelism scheme.
Since we assign each task to a warp, all threads in a warp does the same DFS walk synchronously, 
which avoids most of the branch divergence.
This creates some redundancy, but since most of execution time is spent on set operations, it is still a good tradeoff.

\noindent
\textbf{Counting-only pruning}.
In \cref{sect:compare-gpu}, we do not enable optimization D in \cref{tab:optimizations},
because GraphZero and Pangolin do not support it.
We observe that for those patterns (e.g., {\tt diamond}) enabling this pruning in \sys further improve performance by 6.2$\times$ on average.
Enabling this optimization in Peregrine also improves its performance, as shown in \cref{tab:count}.
However, due to our high efficiency on GPU, 
\sys still outperforms Peregrine by 41.1$\times$ when both enable it.
This again demonstrates the performance superiority of GPU over CPU, no matter what algorithm optimizations are applied.

\noindent
\textbf{Sorting and renaming vertices}. 
For fair comparison, this optimization done by the preprocessor is also not enabled in \cref{sect:compare-gpu}.
Our evaluation shows that this can futher improve \sys performance by 5\% (up to 90\%).
Applying this to GraphZero also helps,
but \sys is still 12$\times$ faster.


\section{Conclusion}

We present \sys, the first multi-GPU GPM framework that supports efficiently mining large graphs and patterns.
For high efficiency, \sys is aware of the input, pattern and architecture to fully unlock the potential of GPM computing on GPUs, which results in a 5$\times$ speedup over the state-of-the-art GPU-based GPM system, Pangolin, on a single GPU.
For scalability, \sys employs a custom task scheduler that can scale GPM computation to multiple GPUs linearly.
For programmability, it automatically enables applicable optimizations and generates CUDA code, which hides away GPU programming complexity, and in turn provides the same easy-to-use programming interface as the state-of-the-art CPU-based GPM frameworks (e.g., Peregrine). 
We also show that \sys on a single V100 GPU is 48$\times$ faster than Peregrine on a 56-core Intel CPU, a free lunch for GPM users.


\section{Acknowledgements}
\hl{This research is funded by Samsung Semiconductor (GRO grants) and
MIT-IBM Watson AI Lab,
and supported by XSEDE allocation TG-CIE-170005 and ASC22045.
We thank Tianhao Huang and OSDI reviewers for their feedback.  
}

\bibliographystyle{plain}
\bibliography{references}

\newpage

\appendix
\section{Artifact Appendix}

\subsection*{Abstract}

This artifact appendix helps the readers reproduce
the main evaluation results of the OSDI’ 22 paper:
Efficient and Scalable \gpm on GPUs.

\subsection*{Scope}

The artifact can be used for evaluating and reproducing the main results of the paper, 
including \cref{tab:tc}, \cref{tab:cl}, \cref{tab:sgl}, \cref{tab:mc}, \cref{tab:fsm} and
\cref{fig:three graphs}, \cref{fig:distr}, \cref{fig:k-clique}, \cref{fig:warp-eff} in \cref{sect:eval}.

\subsection*{Contents}

The artifact evaluation includes all the experiments in the paper.
Details of the experiments are listed \href{https://github.com/chenxuhao/GraphMiner/blob/master/OSDI-experiments-guide.md}{here}:
https://github.com/chenxuhao/GraphMiner/blob/master/OSDI-experiments-guide.md

\subsection*{Hosting}

The source code of this artifact can be found
on \href{https://github.com/chenxuhao/GraphMiner}{GitHub}: https://github.com/chenxuhao/GraphMiner, master branch.

\subsection*{Requirements}

\textbf{Hardware dependencies}

This artifact depends on an NVIDIA V100 GPU.

\noindent
\textbf{Software dependencies}

This artifact requires CUDA toolkit 11.1.1 or greater and GCC 8 or greater.

Details of the dependencies are listed \href{https://github.com/chenxuhao/GraphMiner/blob/master/OSDI-experiments-guide.md}{here}:
https://github.com/chenxuhao/GraphMiner/blob/master/OSDI-experiments-guide.md



\end{document}